\documentclass[aip,
	pop,
	reprint,
	floatfix]{revtex4-2}

\usepackage{graphicx}
\usepackage[left=1in, right=1in,  top=1in, bottom=1in]{geometry}
\usepackage{amsmath}
\usepackage{amssymb}
\usepackage{subfig}
\usepackage{float}
\usepackage{etoolbox}

\usepackage[utf8]{inputenc}
\usepackage[T1]{fontenc}
\usepackage{bm}%

\makeatletter
\def\@email#1#2{%
 \endgroup
 \patchcmd{\titleblock@produce}
  {\frontmatter@RRAPformat}
  {\frontmatter@RRAPformat{\produce@RRAP{*#1\href{mailto:#2}{#2}}}\frontmatter@RRAPformat}%
  {}{}
}%
\makeatother

\begin{document}
\preprint{AIP/123-QED}

\title[]{Machine-learned trends in mirror configurations in the Large Plasma Device}

\author{{Phil Travis\raise0.7ex\hbox{$\ast$}}}%
  \affiliation{Department of Physics and Astronomy, University of California, Los Angeles, CA, USA}
  \email{phil@physics.ucla.edu.}
\author{Jacob Bortnik}%
  \affiliation{Department of Atmospheric and Oceanic Sciences, University of California, Los Angeles, CA, USA}
\author{Troy Carter}%
  \affiliation{Department of Physics and Astronomy, University of California, Los Angeles, CA, USA}
  \affiliation{Oak Ridge National Laboratory, Oak Ridge, TN, USA}

\date{\today}%

\begin{abstract}
This study demonstrates the efficacy of machine learning (ML)-based trend inference using data from the Large Plasma Device (LAPD). The LAPD is a  flexible basic plasma science device with a high discharge repetition rate (0.25-1 Hz) and reproducible plasmas capable of collecting high-spatial-resolution probe measurements. A diverse dataset is collected through random sampling of LAPD operational parameters, including the magnetic field strength and profile, fueling settings, and the discharge voltage. Neural network (NN) ensembles with uncertainty quantification are trained to predict time-averaged ion saturation current ($I_\text{sat}$ — proportional to density and the square root of electron temperature) at any position within the dataset domain. Model-inferred trends, such as the effects of introducing mirrors or changing the discharge voltage, are consistent with current understanding. In addition, axial variation is optimized via comprehensive search over $I_\text{sat}$ predictions. Experimental validation of these optimized machine parameters demonstrate qualitative agreement, with quantitative differences attributable to Langmuir probe variation and cathode conditions. This investigation demonstrates, using ML techniques, a new way of extracting insight from experiments and novel optimization of plasmas. The code and data used in this study are made freely available.

\end{abstract}

\pacs{}%

\maketitle %

\section{Introduction}

Understanding and controlling plasma behavior in fusion devices is necessary for developing efficient fusion reactors for energy production. Because of the complex, high-dimensional parameter space, traditional experimental approaches are often time-consuming and require careful planning. This work explores how machine learning (ML) techniques can accelerate this understanding by studying the effect of machine parameters in a basic magnetized plasma device. Trend inference is this process of relationship discovery. While ML methods, particularly neural networks (NNs), have become increasingly prevalent in fusion research for control and stabilization, their application to systematic trend discovery remains largely unexplored.

Many studies have used ML for profile prediction on a variety of tokamaks, particularly for real-time prediction and control. For example, NNs were used to predict electron density, temperature, and other quantities in DIII-D \cite{abbate_data-driven_2021}, and reservoir NNs have demonstrated the ability to quickly adapt to new scenarios or devices \cite{jalalvand_real-time_2022}. Temporal evolution of parameters has been successfully modeled using recurrent neural networks (RNNs)\cite{char_full_2024} for multiple devices, including the EAST\cite{wan_east_2022} and KSTAR tokamaks\cite{seo_feedforward_2021,seo_development_2022}. These predictions enabled training of a reinforcement learning-based controller\cite{seo_feedforward_2021,seo_development_2022}. In addition, a decision tree-based controller was trained to maximize $\beta_N$ while avoiding tearing instabilities\cite{fu_machine_2020} on DIII-D. Electron temperature profiles have also been predicted using dense NNs on the J-TEXT tokamak \cite{dong_machine_2021}.

A parallel focus has been on instability prediction and mitigation in tokamaks, particularly of disruptions. Notable achievements in disruption prediction include RNN-based disruption prediction \cite{kates-harbeck_predicting_2019} and random forest approaches\cite{rea_real-time_2019}, with a comprehensive review available by Vega et al \cite{vega_disruption_2022}. Recent work has extended to active control, such as the mitigation of tearing instabilities in DIII-D using reinforcement learning \cite{seo_avoiding_2024}. 

While ML has proven effective for prediction and control tasks, inferring trends using data-driven methods has been relatively uncommon. Notable exceptions include finding scaling laws on the JET tokamak\cite{murari_investigating_2020} via classical ML techniques and the development of the Maris density limit\cite{maris_correlation_2024} which outperforms other common scalings (including the Greenwald density limit) in predictive capability.

The use of machine learning and Bayesian inference in fusion research has been recently reviewed by Pavone et al.\cite{pavone_machine_2023}

Outside of magnetized plasmas, the laser plasma community has embraced ML techniques for various applications, enumerated in a review by Dopp et al.\cite{dopp_data-driven_2023}. 
Data-driven plasma science in general has been reviewed by Anirudh et al.\cite{anirudh_2022_2023}
Notably, a similar quasi-random method (Sobol sequences) was used to collect a spectroscopy dataset on a plasma processing device over diverse machine settings \cite{daly_data-driven_2023}. This process is similar to what is performed in our work here, but a generative variational autoencoder was instead trained to be used as an empirical surrogate model.

This work advances data-driven methods in plasma physics by taking these methods one step further: instead of learning a model for particular task (e.g., disruption prediction or profile prediction), we infer learned trends directly from the model itself. 

The goal of this study is to develop a data-driven model that can provide insight into the effect of machine parameters on plasmas produced in Large Plasma Device (LAPD) in lieu of a theoretical model. In contrast with tokamaks and other fusion devices, the LAPD is particularly well-suited for ML data collection because of its flexibility and high repetition rate. We demonstrate the capability to infer trends in a particular diagnostic signal, the time-averaged ion saturation current ($I_\text{sat}$), for any mirror (or anti-mirror) field geometry in a variety of machine configurations. Langmuir probes are commonly used to measure density, temperature, and potential in virtually all plasma devices in low-temperature (less than 10s of an eV) regimes. The $I_\text{sat}$ signal in particular is almost always used in the LAPD for calculating local plasma density.

This study performs two firsts in magnetized plasma research: using NNs to directly infer trends and collecting data efficiently with partially-randomized machine parameters. We also demonstrate optimizing LAPD plasmas given any cost function by minimizing axial variation in $I_\text{sat}$. This global optimization is only possible using ML techniques. This work demonstrates the usefulness of a pure ML approach to modeling device operation and shows how this model can be exploited. We encourage existing ML projects and experiments to consider this approach if possible. Acquiring sufficiently diverse datasets may require assuming some risk because diverse data, such as discharges from randomly sampled machine settings, may not be amenable to conventional analysis techniques.

All the processed data used for training the models in this study are freely available\cite{phil_travis_2025_15007853} (see Appendix \ref{app:repo}). Other devices have also made data publicly available. Namely, data for H-mode confinement scaling has been available since 2008\cite{roach_2008_2008}, and more recently some MAST\cite{jackson_fair-mast_2024} and all LHD\cite{lhd_data} data are now publicly available.  

This paper is organized as follows: Section \ref{sec:data} discusses the LAPD and the data acquisition methodology. Sections \ref{sec:model_dev} and \ref{sec:uncertainty} detail development of the model and uncertainty quantification. Section \ref{sec:validation} presents model validation, followed by trends in discharge voltage and gas puff duration in Section \ref{sec:trends}. Section \ref{sec:optimization} demonstrates optimization of the axial $I_\text{sat}$ profile, with the discussion and conclusion in sections \ref{sec:discussion} and \ref{sec:conclusion}. 

\section{Data collection and processing}
\label{sec:data}

\subsection{The Large Plasma Device (LAPD)}

\begin{figure*}
	\centering
	\includegraphics[width=\textwidth]{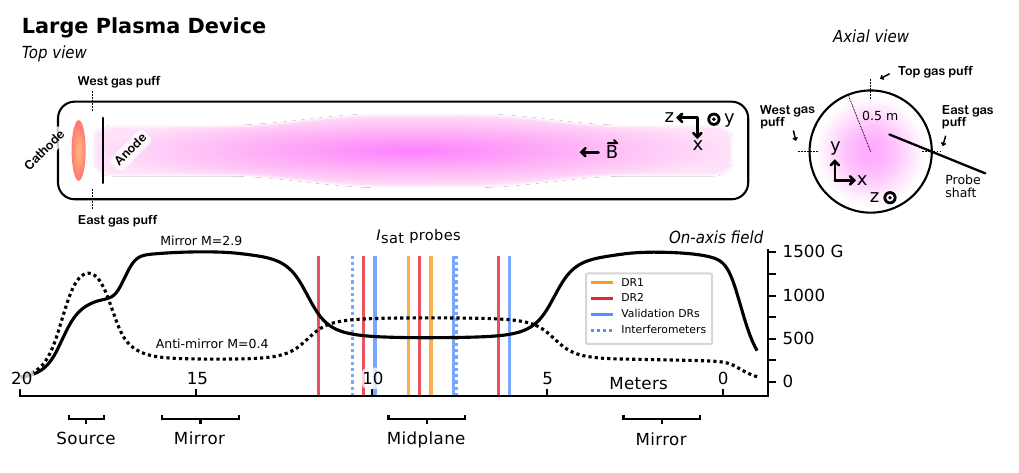}
	\caption[size=12]{\label{fig:LAPD_coords}A cartoon of the Large Plasma Device, the coordinate system used, examples of a mirror and anti-mirror magnetic field configuration, and probe locations used in this study. The source, mirror, and midplane regions are labeled; the three fields were programmed independently.}
\end{figure*}

The Large Plasma Device (LAPD)\cite{gekelman_upgraded_2016,qian_design_2023} is a basic plasma science device located at the University of California, Los Angeles. The LAPD produces up to 18m long, 1m diameter plasmas  with densities up to $3 \times 10^{13}$ cm$^{-3}$ and temperatures up to 20 eV, though typical operation yields temperatures around 5 eV. Probes can sample virtually any point in this plasma through  unique ball valves placed every 32 cm along the length of the device, enabling the collection of time series data with high spatial resolution. The discharge repetition rate is configurable between 0.25 and 1 Hz. Additionally, the LAPD has 13 independently controllable magnet power supplies to shape the geometry of the axial magnetic field. The discharge is formed by 35 cm diameter lanthanum hexaboride (LaB6) cathode\cite{qian_design_2023} and 72 cm molybdenum anode 0.5m away (-z direction) at the southern end (+z) of the device. A cartoon of the LAPD and relevant coordinate system can be seen in Fig. \ref{fig:LAPD_coords}. 

The LAPD has many experimental control parameters for various physics studies. While the device can accommodate various insertable components, this study focuses on the parameters fundamental to the operation of the main cathode. Specifically, half way between the cathode and anode are three gas puff valves: East, West, and top. The aperture, duration, and triggering of these valves has a large impact on plasma formation. A static gas fill system also exists but it is not used in this study. The cathode-anode voltage (and consequently, discharge power) strongly influences plasma density and temperature downstream of the source. Additionally, the magnetic field configuration substantially shapes the plasma column. One crucial variable not considered in this study is the cathode temperature, as its adjustment and equilibration requires many hours, limiting dataset diversity. This combination of diagnostic coverage, high repetition rate, and extensive configurability renders the LAPD particularly suitable for machine learning studies. 

\subsection{Data collection and processing of $I_\text{sat}$ signals}

The ion saturation current, denoted as $I_\text{sat}$, is obtained by applying a sufficiently negatively bias to a Langmuir probe to ensure the exclusive collection of ions. This collected current is proportional to $S n_e \sqrt{T_e}$, where $n_e$ and $T_e$ are the electron density and temperature, and $S$ is the effective probe collection area. To account for differences in probe tip geometry, the $I_\text{sat}$ values are normalized to area. Under typical conditions, an $I_\text{sat}$ value of 1 mA/mm$^2$ corresponds to $n_e \approx 1\text{-}2\times 10^{12}$ cm$^{-3}$ for a $T_e$ from 4 to 1 eV.

Data collection was conducted in two campaigns separated by 14 months. The initial run set is designated as \texttt{DR1} and the subsequent run set as \texttt{DR2}. These run sets are further broken down into \em dataruns \em which are series of discharges (``shots'') with identical operational machine parameters. A total of 67 dataruns were collected over both campaigns. 

\begin{figure}
	\centering
	\includegraphics[width=\columnwidth]{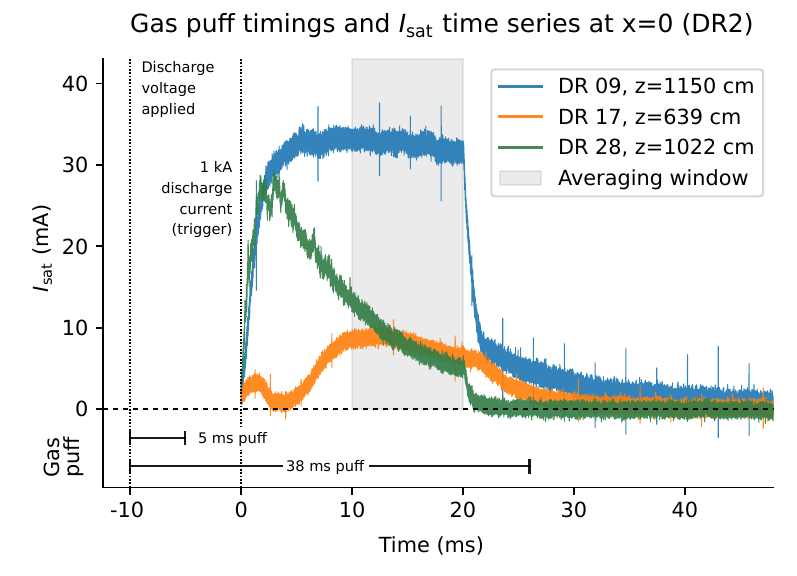}
	\caption{\label{fig:PP1_time-series-example}Gas puff timings and example $I_\text{sat}$ time series at three different z-axis locations from three different dataruns. Note that some discharges do not achieve steady state in $I_\text{sat}$. }
\end{figure}

$I_\text{sat}$ measurements were averaged over 10 to 20 ms to exclude plasma ramp-up and fluctuations. Example $I_\text{sat}$ probe data can be seen in fig. \ref{fig:PP1_time-series-example} along side gas puff timings. Profile evolution is not studied to minimize computational requirements. $I_\text{sat}$ characteristics vary significantly between axial (z) position machine parameters. For $I_\text{sat}$ measurements on the same probe as a Langmuir sweep (\texttt{DR2} port 26, z=863 cm), the averaging process excludes the sweep period with an additional 40 $\mu$s buffer. $I_\text{sat}$ measurements in \texttt{DR1} that saturated either the isolation amplifier or digitizer are excluded from the dataset. Only 484 shots were removed out of $\approx$132,000, so the impact on the aggregate dataset is minimal. 

The LAPD control parameters varied in this study were the source field, mirror field, midplane field, gas puff valve voltage, gas puff duration, and discharge voltage. The magnetic field regions are labeled in Fig. \ref{fig:LAPD_coords} and effectively control the width of the plasma relative to the cathode in their respective regions. The gas puff voltage governs gas flow rate into the chamber, though this relationship is not yet quantified, and the gas puff duration defines the piezo valve activation period. The discharge voltage is applied across the cathode and anode 10 ms after gas puff initiation. While discharge voltage correlates to discharge current (and thus power), the current depends on the machine configuration and downstream conditions and cannot be predetermined.

These machine parameters -- with the exception of gas puff duration -- were randomly sampled via Latin-hypercube sampling (LHS) for 44 of the dataruns. Data were then collected with these settings. Gas puff duration was reduced for the last seven runs to 20, 10, or 5 ms (see fig. \ref{fig:PP1_time-series-example} for timings relative to $I_\text{sat}$ signals). The breakdown of each setting in the dataset is given in appendix \ref{sec:app_bias}, Table \ref{tab:data_frac}. The top gas puff valve was used for only the first nine dataruns of \texttt{DR2} because of equipment issues. 

$I_\text{sat}$ measurements were acquired along y=0 lines (51 dataruns total) or x-y grids (16 dataruns total) with spatial resolutions varying between 1.5 to 2 cm. The fixed axial locations of the probes were 895 cm and 831 in \texttt{DR1} and 1150, 1022, 863, and 639 cm for \texttt{DR2} (Fig. \ref{fig:LAPD_coords}). Six shots were recorded at each position except for the first four dataruns in \texttt{DR1} with five shots each. While $I_\text{sat}$ exhibits a small degree of shot-to-shot variation, the present model only learns the expected value, leaving distributional modeling to future generative approaches.

\subsection{Dataset distribution and bias}

Neural network inputs comprise 12 variables: source field, mirror field, midplane field, gas puff voltage, discharge voltage, gas puff duration, probe coordinates (x, y, z), probe rotation, run set identifier, and top gas puff flag. These variables can be interpreted as six control parameters, four probe coordinates, and two flags. These inputs are mean-centered and normalized to the peak-to-peak value with no outliers in the dataset. The baseline models trained in section \ref{sec:baselines} did not contain the run set identifier or top gas puff flag.

Measurements over an x-y plane, constituting $\approx 64\%$ of all shots, are predominantly acquired overnight for maximal machine utilization. These longer dataruns lead to particular machine configurations being overrepresented in the dataset. 

The dataset predominantly contains gas puff durations of 38 ms. Only 6 runs in the training set have gas puff durations less than 38 ms: three have 5 ms and three have 10 ms, each having mirror ratios 1, 3, and 6 but otherwise identical configurations in an attempt to see mirror-related interchange instabilities in higher-temperature, lower-collisionality regimes. The 20 ms gas puff duration case is in the test set. This sampling bias towards the 38 ms gas puff duration suggests poor model performance is to be expected in shorter gas puff regimes. The top gas puff valve was operational for only the first nine runs of \texttt{DR2}.

The $I_\text{sat}$ distribution is dissimilar between \texttt{DR1} and \texttt{DR2}: \texttt{DR1} appears to have a more uniform distribution. Combining the two datasets results in many $I_\text{sat}$ examples less than 2 mA/mm$^2$ and a sharp decrease in number of examples above 10 mA/mm$^2$. Thus, we expect the model to perform better for smaller $I_\text{sat}$ values than larger ones. Data bias is further discussed in appendix \ref{sec:app_bias}.

\section{Model development, training}
\label{sec:model_dev}

\subsection{Training details}

For initial experiments in training the model, a mean-squared error (MSE) loss is used:
\begin{equation}
	\mathcal{L}_\text{MSE}=\frac{1}{m}\sum_{i=1}^m \left(f\left(x_i\right) - y_i \right)^2
\end{equation} 
where $x_i$ represents the input vector for the $i$th example, $y_i$ the target measurement, $m$ the batch size, and $f$ the NN. During training, overfitting was assessed via the validation set MSE with a traditional 80-20 train-validation random split. Unless stated otherwise, a dense neural network, 4 hidden layers deep and 256 units wide ($\approx$200,000 parameters), was trained with AdamW using a learning rate of $3\times10^{-4}$. Leaky ReLU activations (the nonlinearities in the NN) and adaptive gradient clipping\cite{seetharaman_autoclip_2020} (cutting gradients norms above the 90th percentile of recent norms) were used to mitigate vanishing gradients and mitigate exploding gradients, respectively. The models were evaluated after training concluded at 500 epochs.

\subsection{Baselines for mean-squared error}
\label{sec:baselines}

A model was first trained with zeroed-out inputs as a baseline and to validate the data pipeline. This model effectively has only a single, learnable bias parameter at the input. This process yields a validation loss (simply MSE in this case) of 0.036. A linear model, though incapable of reasonably fitting the data, was trained as a performance baseline and to validate the data pipeline. This baseline linear model reaches a training and validation loss (MSE) of around 0.014. These initial models used \texttt{tanh} activations , though the impact of using a different activation function is minimal in these cases.
A summary of these baselines is seen at the top of Table \ref{tab:loss_summary}.

\begin{table}
	\small
	\centering
	\caption{Summary of test set losses for different training data and ensembles}
	\label{tab:loss_summary}
	\begin{tabular}{l l l}
		Model & MSE $\times 10^{-3}$ \\
		\hline
		Zeroed-input & 36  (validation) \\
		Linear model & 14 (validation) \\
		\hline
		9 dataruns & 7.0\\
		19 dataruns & 6.9 \\
		29 dataruns & 4.2 \\
		39 dataruns & 4.1 \\ 
		49 dataruns & 3.4 \\
		\texttt{DR1} only & $6.4$ \\
		\texttt{DR2} only & $5.4$ \\
		Full set, large model & $2.8$ \\
		Full set average & $3.6 \pm 0.56$ \\
		Full set ensemble & $2.9 \pm 1.1$ \\
		\hline
		``Run set'' flag ensemble & $1.9 \pm 0.64$ \\
		``Top gas puff'' flag & 1.8 \\
		
	\end{tabular}
\end{table}

\subsection{Effects of training set and model sizes}
 
To study the effects of reduced diversity, the number of unique dataruns in the training set was systematically reduced while evaluating on a fixed test set. The test set loss monotonically increased with this decrease in datarun count. Part of this decrease may be caused by a simple reduction in training set size. In addition, models were individually trained and evaluated on \texttt{DR1} only or  \texttt{DR2} only. When evaluated on the left-out run set, the test set losses were high, near or above the zero-input baseline of $3.6 \times 10^{-2}$. This result suggests that both run sets contain significant information missing in the other, and training on both provides beneficial information on the structure of the $I_\text{sat}$ measurement despite different probe calibrations and cathode state.

A larger model, consisting of a 12-deep 2048-wide dense network, was trained on the full training dataset, evaluated at 30 epochs. This larger model yielded a test MSE of $2.8 \times 10^{-3}$, indicating that these NNs are behaving as expected. Longer training or larger models may yield better test set results, but will likely not come close to the training and validation losses which are on the order of $10^{-5}$. Combined models with differing initializations (an ensemble), were trained to measure the MSE variance over model parameters which was about 16\%. When the $I_\text{sat}$ predictions were averaged, the test set MSE was $2.9 \pm 1.1 \times 10^{-3}$, achieving the best performance for that model size. These test set losses are also seen in Table \ref{tab:loss_summary}.

\subsection{Improving performance with machine state flags}

Data from \texttt{DR1} and \texttt{DR2} were collected 14 months apart leading to differing machine states. In \texttt{DR1}, only one turbo pump was operating leading to much higher neutral pressures than in the \texttt{DR2} run set. A new parameter (mean-centered and scaled) was added to the inputs to distinguish between these two run sets. All the predictions in this study use the \texttt{DR2} run set flag (a value of 1.0) because turning off the turbopumps is not a commonly desired mode of operating the LAPD. The inclusion of this parameter also provides the model the ability to distinguish between the probe calibration differences between \texttt{DR1} and \texttt{DR2}. An ensemble prediction with this run set flag brings the test set MSE down to $1.9 \times 10^{-3}$.  

A flag indicating when the top gas puff valve was enabled in \texttt{DR2} was also added to all training data, allowing the model to further distinguish between different fueling cases. The addition of this flag incrementally improved test set MSE to $1.8 \times 10^{-3}$. The effect on MSE on the inclusion of these new parameters is compared to the performance of other models in Table \ref{tab:loss_summary}. 

\subsection{Learning rate scheduling}
Modifying the learning rate over time (scheduling) is known to improve model learning. The following schedules were compared: constant learning rate ($\gamma = 3 \times 10^{-4}$), $\gamma \propto \text{epoch}^{-1}$, $\gamma \propto \exp{(-\text{epoch})}$, and $\gamma \propto \text{epoch}^{-1/2}$. The epoch is the training step divided by the number of batches in one epoch, so ``epoch'' in this case takes on a floating-point value. $\gamma \propto \text{epoch}^{-1}$ appears to give the best test set loss by a test MSE difference of $1 \times 10^{-4}$, and any schedule beats a constant learning rate by a difference of $2-4 \times 10^{-4}$.

\section{Uncertainty quantification}
\label{sec:uncertainty}

\subsection{$\beta$-NLL loss}

Instead of predicting a single point, the model can predict a mean $\mu$ and variance $\sigma^2$ using the negative-log likelihood (NLL) loss \cite{nix_estimating_1994,lakshminarayanan_simple_2017} by assuming a Gaussian likelihood. An adaptive scaling factor $\text{StopGrad}(\sigma_i^{2\beta})$ is introduced that can be interpreted as an interpolation between an MSE loss and Gaussian NLL loss, yielding the $\beta$-NLL loss:

	\begin{multline}	
	\mathcal{L}_{\beta-\text{NLL}} = \frac{1}{2}\left( \log{\sigma^2_i(\mathbf{x}_n)} +\frac{\left(\mu_i(\mathbf{x}_n) - y_n\right)^2}{\sigma^2_i (\mathbf{x}_n)} \right) \\ \text{StopGrad}\left(\sigma_i^{2\beta}\right)
	\label{eq:loss_beta-NLL}
	\end{multline}

for example $n$ and model $i$, with an implicit expectation over training examples. $\beta=0$ yields the original Gaussian NLL loss function and $\beta=1$ yields the MSE loss function. This factor improves MSE performance by scaling via an effective learning rate for each example (which necessitates the \texttt{StopGrad} operation) \cite{seitzer_pitfalls_2022}, and improves both aleatoric and epistemic uncertainty quantification \cite{valdenegro-toro_deeper_2022}. $\beta=0.5$ was used by default in this study. This $\beta$-NLL loss function also improved training stability.

This NLL-like loss assumes the prediction -- the likelihood of $y$ given input $\mathbf{x}$: $p(y|\mathbf{x})$ -- follows a Gaussian distribution. Treating each prediction as an independent random variable (considering each model in the ensemble is sampled from some weight distribution $\theta \sim p(\theta | \mathbf{x}, y)$) and finding the mean of the random variables yields a normal distribution with mean $\mu_* (\mathbf{x}) = \langle \mu_i(\mathbf{x}) \rangle $ and variance $\sigma^2_* = \langle \sigma^2_i(\mathbf{x}) + \mu^2_i(\mathbf{x}) \rangle - \mu^2_* (\mathbf{x})$ where $\langle \rangle$ indicates an average over the ensemble.

The ensemble predictive uncertainty can be broken down into the aleatoric and epistemic components \cite{valdenegro-toro_deeper_2022}: the aleatoric uncertainty is $\langle \sigma^2_i (\mathbf{x}) \rangle$ and the epistemic uncertainty is $\langle \mu^2_i (\mathbf{x}) \rangle - \mu^2_* (\mathbf{x}) = \text{Var}\lbrack\mu_i (\mathbf{x}) \rbrack$. The intuition behind these uncertainties is that the random fluctuations in the recorded data are captured in the variance of a single network, $\sigma^2_i$. If the choice of model parameters were significant, we would expect the predicted mean for a single model, $\mu_i$, to fluctuate as captured by $\text{Var}\lbrack\mu_i (\mathbf{x}) \rbrack$.

\subsection{Cross-validation MSE}

For cross-validation, multiple train-test set pairs were created. Test set 0 comprises deliberately chosen dataruns to encompass a diverse set of machine settings and probe movements. The other six datasets were  compiled with randomly chosen dataruns (without replacement) while keeping the number of dataruns from \texttt{DR1} and \texttt{DR2} equal. Seven model ensembles (5 NNs per ensemble -- 35 NNs total) were trained to evaluate the effect of test set choice on model MSE. The median ensemble test set MSE for these seven sets was $2.13 \times 10^{-3}$ with a mean of $3.6 \times 10^{-3}$. The handpicked dataset had an ensemble test set MSE of $1.85 \times 10^{-3}$, indicating that the choice of dataruns was adequately representative. This median MSE will be used to estimate model prediction error in addition to uncertainty quantification. This cross-validation also provides an error estimate if the models were to be trained on \emph{all} dataruns. Ensembles always out-performed the average error of single-model predictions.

All validation set MSEs fall between $1$ and $6 \times 10^{-5}$, with the average training MSE falling within that range as well. These MSEs indicate that the model is able to fit the training data to a high degree of accuracy regardless of which dataruns are held out. The loss and MSE curves over training epochs can be seen in the appendix in fig. \ref{fig:beta-NLL_loss-mse}.

\subsection{Model calibration via weight decay}

The predicted uncertainty may not provide an accurate range of $I_\text{sat}$ values when compared to the measured value. Calibrating the model means changing the predicted uncertainty range so that the measured values fall within that range according to some distribution, such as a Gaussian in this case. One of the ways assessing this calibration is by the z-score of predictions, where $z_n= \left(x_n - \mu_n \right) / \sigma_n(x_n)$ for example $x_n$, predicted mean $\mu_n$, and standard deviation $\sigma_n$. Perfect model calibration would lead to identical z-score distribution $\mathcal{N}(\mu=0, \sigma=1)$ for the training a test sets. When evaluated on the training set, the distribution should be a Gaussian with a standard deviation of 1.

\begin{figure}
	\centering
	\includegraphics[width=\columnwidth]{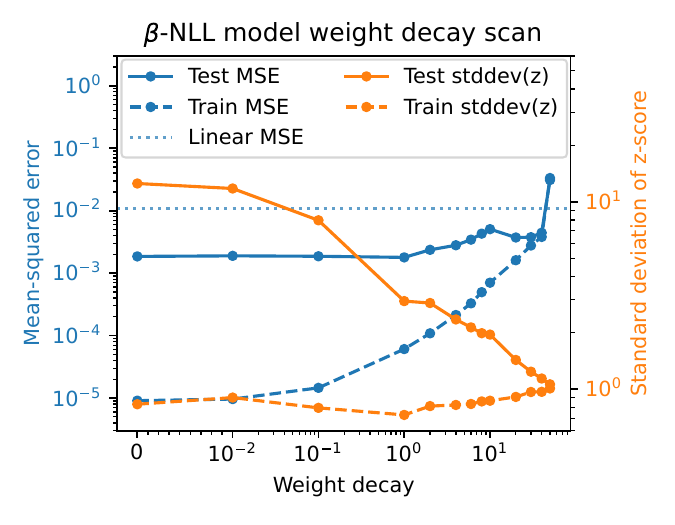}
	\caption[size=12]{\label{fig:beta-NLL_wd_model_performance}Model performance and calibration for different weight decays. Highly biased models are better calibrated, but come at great expense of mean prediction error. At the weight decay value of 50, the model has worse error than a linear model. Note the linear scale below $10^{-2}$.}
\end{figure}

Increased weight decay can lead to better model calibration \cite{guo_calibration_2017}. Weight decay penalizes large parameter values by adding the L2 norm of model weights to the loss. Model ensembles were trained with weight decay coefficients between 0 and 50 to determine the best calibrated model determined by the distribution of z-scores of the training and test sets. The results of this weight decay scan are seen in Fig. \ref{fig:beta-NLL_wd_model_performance}. Increasing the weight decay increases the test MSE and decreases its z-score standard deviation. This large standard deviation is caused by outliers. Excluding z-scores magnitudes above 10, or 4.4\% of the test set, yields a standard deviation of 2.53. This long tail indicates that the distribution of predictions on the test set is not Gaussian. Nonetheless, the trend remains that increasing weight decay leads to smaller test set z-score standard deviations. However, the test set MSE increases after a weight decay of 1. This increase in test MSE implies that the model is making less accurate predictions but is better calibrated. Highly biased models are better calibrated, but come at great expense of mean prediction error. At the weight decay value of 50, the model has worse error than a linear model. Despite the attempts using weight decay, the model never becomes well-calibrated: the predicted uncertainty is always too low by a factor of 2 to 5.

\begin{figure*}
	\centering
	\includegraphics[width=360pt]{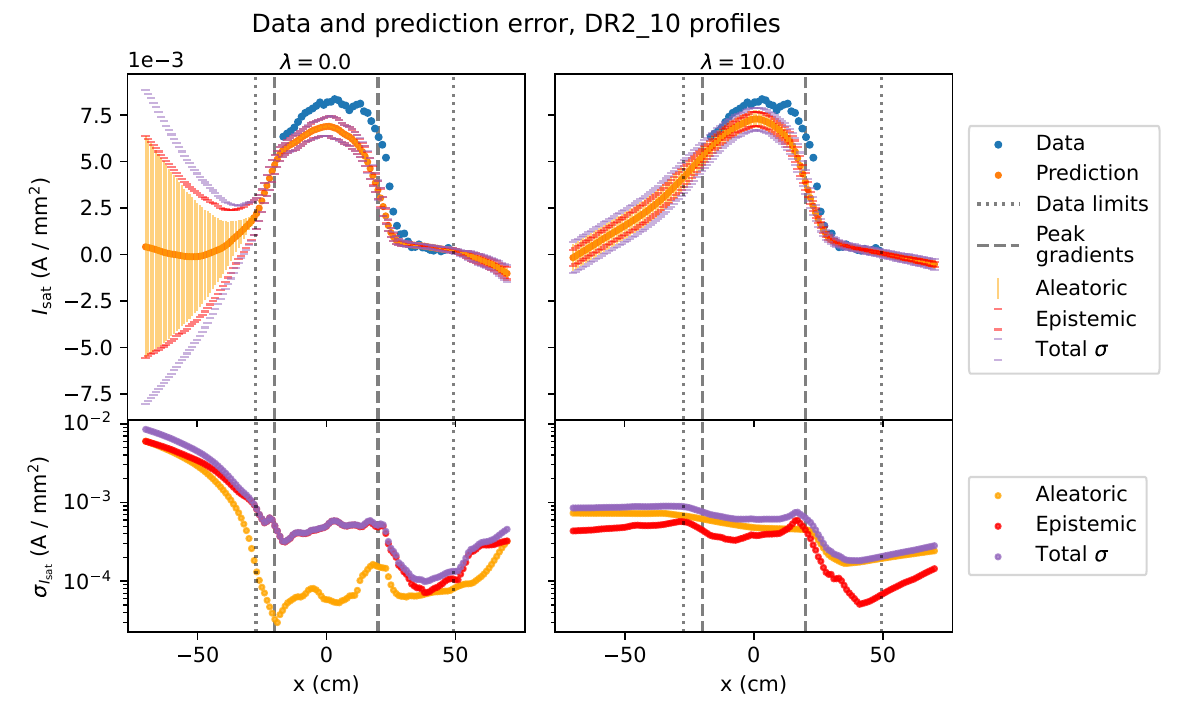}
	\caption[size=12]{\label{fig:extrapolation-profile-var_two}Model extrapolation performance (top plots) with uncertainty (bottom plots) for a model ensemble trained on a $\beta$-NLL loss function. \texttt{DR2} run 10 was chosen as an illustrative example. The \em relative \em uncertainty appears to be more useful when zero weight decay ($\lambda = 0$, left) is used: the uncertainty increases when the model is predicting outside its training data along the x-axis.}
\end{figure*}

Despite the better calibration, the uncertainty predicted by a model with a large weight decay is decidedly worse: the uncertainty is similar across an entire profile, and when projected beyond the training data, the total uncertainty remains largely constant as seen in Fig. \ref{fig:extrapolation-profile-var_two}. The zero weight decay model exhibits relatively increasing uncertainty beyond the bounds of the training data. Although not well-calibrated, this uncertainty can provide a hint of where the model lacks confidence relative to other predictions, even though the uncertainty is much less than it should be.

\section{Evaluating model performance}
\label{sec:validation}

Model performance is evaluated in three ways by comparing against intuition from geometry, an absolute measurement, and extrapolated machine conditions.

\subsection{Checking geometrical intuition}

Assuming magnetic flux conservation, we know that modifying the mirror geometry can control the effective width of the plasma. One way to check that the model is learning appropriate trends is to check with this intuition. If the magnetic field at the source is not equal to the field at the probe, the probe will see the plasma expanded (or contracted) by roughly a factor of $\sqrt{B_\text{probe}/B_{source}}$. The cathode is about 35 cm in diameter, so a magnetic field ratio of 3 would give produce a plasma approximately 60 cm in diameter. All the probes used in this study are in or very close to the zero-curvature midplane region of a mirror.

\begin{figure}
	\centering
	\includegraphics[width=\columnwidth]{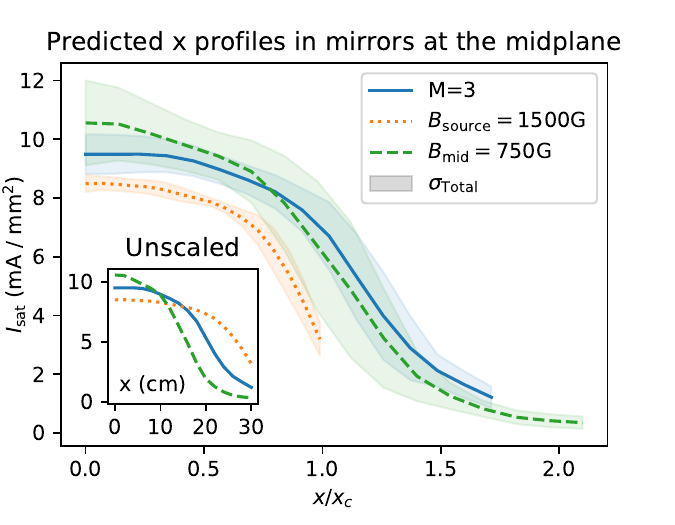}
	\caption[size=12]{\label{fig:changing-B-field_M=3_x-prof}Plot of various mirror configurations scaled to the cathode radius $x_c=17.5$ cm at the midplane (z=790 cm). When scaled according to the expected magnetic expansion, the profiles generally agree. The smaller the plasma diameter (and thus smaller volume), the higher the peak in $I_\text{sat}$ at the core, as expected. }
\end{figure}

To check this intuition, the model is given the following inputs: $B_\text{source}$=500 G, $B_\text{mirror}$=1500 G, $B_\text{midplane}$=500 G, discharge voltage=110 V, gas puff voltage=70 V, gas puff duration=38 ms, run set flag=\texttt{DR2} and top gas puff=off. The discharge voltage and gas puffing parameters were arbitrarily chosen. The x coordinate is scanned from 0 to 30 cm, and the z coordinate from 640 to 1140 cm. This discharge is then modified by separately changing $B_\text{source}$ to 1500 G and $B_\text{midplane}$ to 750 G (M=1.5). %
The x profiles at the midplane (z=790 cm) of the reference M=3 prediction, source field change, and midplane field change, all scaled to cathode radius, can be seen in Fig. \ref{fig:changing-B-field_M=3_x-prof}. Changing the source field to 1500 G increases the $I_\text{sat}$ towards the edge of the plasma, as expected. When the midplane field is increased, the $I_\text{sat}$ values decrease at the edge and increase at the core (x=0 cm), implying a thinner plasma column and is consistent with previously measured behavior. When only the mirror field is modified (not shown), the strongest effect on $I_\text{sat}$ is on or near x=0 cm, and the plasma column width does not appear to appreciably change. 

\subsection{Directly comparing prediction to measurement}

$I_\text{sat}$ measurements were taken with the following LAPD machine settings: $B_\text{source}$=1250 G, $B_\text{mirror}$=500 G, $B_\text{midplane}$=1500 G, discharge voltage=90 V, gas puff voltage=90 V, gas puff duration=38 ms, run set flag=\texttt{DR2} and top gas puff=off. These settings were from a previous discharge optimization attempt. The probes utilized were the permanently-mounted 45$^\circ$ probe drives. These probes were known to have identical effective areas relative to each other from the previous experiment and from analyzing the discharge rampup.

\begin{figure}
	\centering
	\includegraphics[width=\columnwidth]{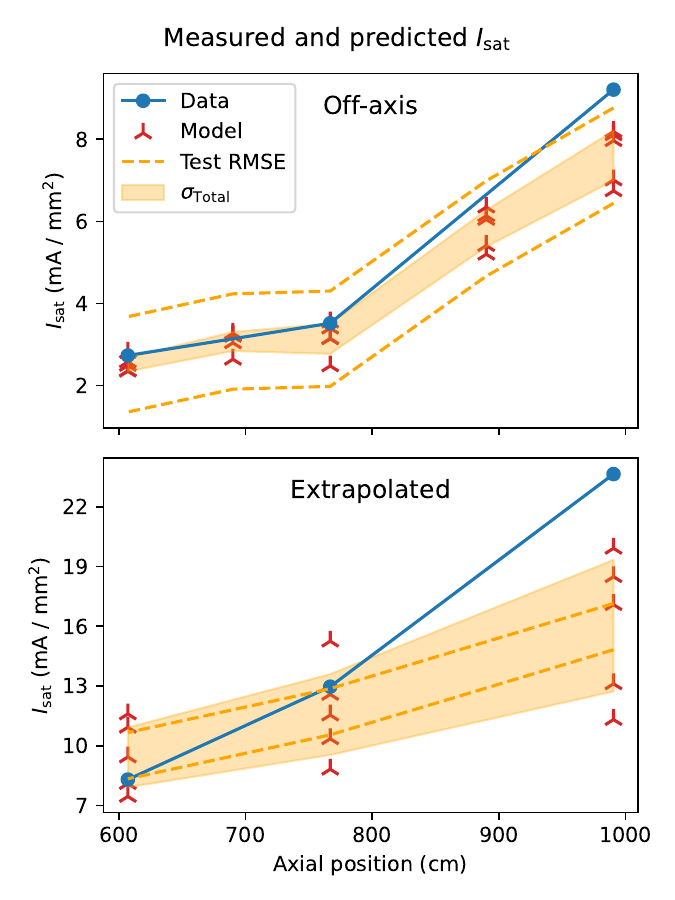}
	\caption[size=12]{\label{fig:measured-vs-predicted}Top: data collected at off-axis positions around x $=9.75$ cm and y $=-8.4$ cm are compared with predictions from the machine learning model at the same points in addition to two interpolating predictions. The model predicts the trend well, but underestimates $I_\text{sat}$ in general. The shaded orange region is the total model uncertainty ($\sigma = \sqrt{\text{Var}}$). Bottom: Measured vs predicted $I_\text{sat}$ values for an odd machine configuration with $B_\text{source}$=822 G and discharge voltage=160 V. The training data only covers discharge voltages up to 150 V The machine was also in an odd discharge state, so it's no surprise that the predicted uncertainty bounds are very large (even greater than the test set RMSE value) and that accuracy suffers.}
\end{figure}

Because of data acquisition issues, only a single useful shot was collected at a nominal -45$^\circ$ angle (relative to the x-axis) 10 cm past the center (x=0 cm, y=0 cm) of the plasma on three probes at z-positions of 990, 767, and 607 cm (ports 22, 29, and 34, respectively). The probe drives were slightly uncentered, leading to the real coordinates of the probes to be around x $=9.75$ cm and y $=-8.4$ cm. Note that the model can predict anywhere in LAPD bounded by the training data, so off-axis measurements are not an issue.
The resulting predictions using these coordinates and machine conditions can be seen in Fig. \ref{fig:measured-vs-predicted}. 
The model reproduces the axial trend well, but slightly underestimates $I_\text{sat}$ on an absolute basis. However, given the lack of absolute $I_\text{sat}$ calibration and variable machine state, the agreement of the absolute $I_\text{sat}$ values may be coincidental. Nevertheless, the trend exhibited by this validation study match the predicted trend and increase our trust in model predictions. 

An additional validation datarun was performed. For this run, the discharge voltage was increased to 160 V, and the source field changed to 822 G. The training data contains discharge voltages up to 150 V, so this case tests the extrapolation capabilities of the model. The comparison of model predictions and the measured data can be seen in Fig. \ref{fig:measured-vs-predicted}. As stated earlier, the absolute uncertainty provided of the model is not calibrated. However, note that the level of uncertainty provided by the model, as well as the large spread in model predictions, are much greater than seen in the interpolation regime (Fig. \ref{fig:measured-vs-predicted}) and eclipses the cross-validated test set root mean squared error (RMSE). We conclude that this model has good interpolation capabilities, but extrapolation -- as with any model -- remains difficult.

\section{Inferring trends}
\label{sec:trends}

\begin{figure}
	\centering
	\includegraphics[width=\columnwidth]{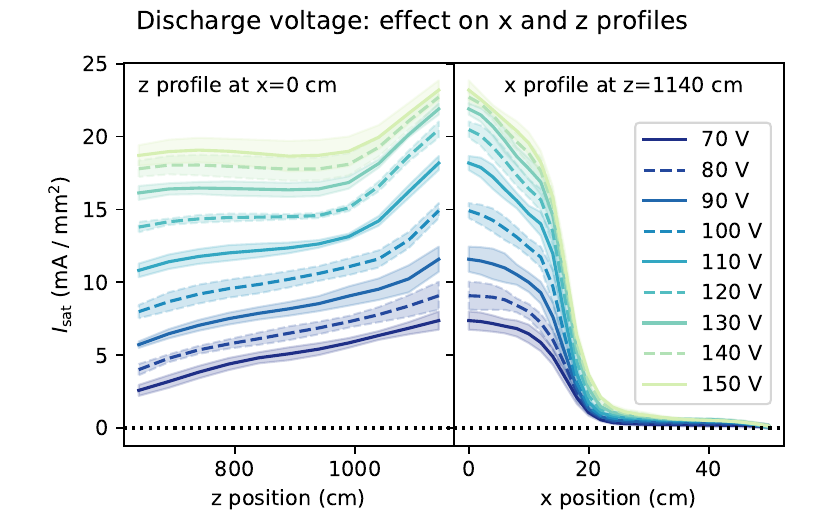}
	\caption[size=12]{\label{fig:discharge_voltage_effect}The z profile at x=0 cm and x profile at z=1140 cm for different discharge voltages. The $I_\text{sat}$ decreases with increasing voltage, and the error (filled regions) stays roughly the same, but in general increase slightly towards the cathode and at higher discharge voltages.}
\end{figure}

A systematic study of the impact of discharge voltages on $I_\text{sat}$ profiles has not been conducted using conventional techniques. Collecting both z- and x-axis profiles over a wide range of discharge voltages would take a considerable amount of time, mostly from the requirement to dismount and reattach the probes and probe drives along the length of the LAPD. This study has now been performed using the learned model, circumventing these time-consuming challenges. Model input parameters were chosen to be common, reasonable values: 1 kG flat field, 80 V gas puff, 38 ms gas puff duration, run set=\texttt{DR2}, and top gas puff off. The 38 ms puff is used in these predictions because it is the most common gas puff duration in the training set -- the model is biased in favor of this gas puff setting. The results of changing the discharge voltage only can be seen in fig \ref{fig:discharge_voltage_effect}. Notably, the $I_\text{sat}$ increases across both axes. Steeper axial gradients are seen with lower discharge voltages, but peaked x-profiles are seen at higher discharge voltages. The area closer to the source region (+z direction) appears to have a steep drop but flatter profiles down the length of the machine. 

 Unfortunately the discharge current was not included as an output in the training set. Otherwise the effect of changes in discharge power, rather than simply voltage, could be computed. The discharge current -- and thus discharge power -- is set by cathode condition, cathode heater settings, and the downstream machine configuration, and thus cannot be set to a desired value easily before the discharge. Discharge voltage, however, can remain fixed.

\begin{figure}
	\centering
	\includegraphics[width=\columnwidth]{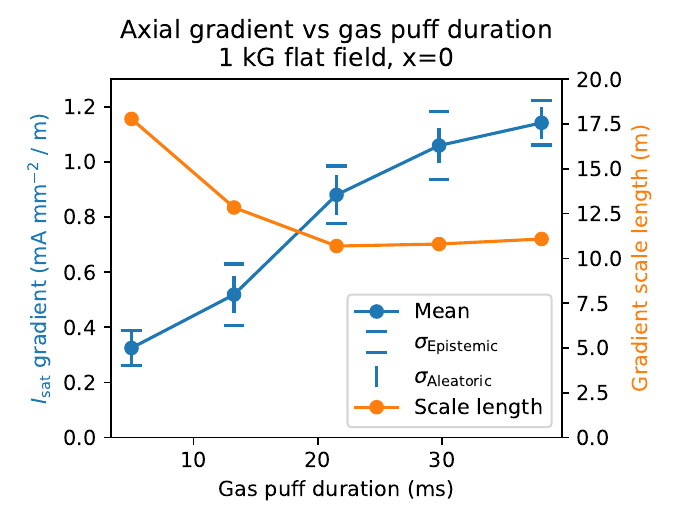}
	\caption[size=12]{\label{fig:axial-grad_gas-puff.pdf}ML prediction: mean axial gradients decrease with decreased gas puff duration. Five durations are plotted between 5 and 38 ms (which are the bounds of the training data), evenly spaced. The gradient scale length also increases, indicating that the gradient change was not just from a decrease in the bulk $I_\text{sat}$.}
\end{figure}

Of particular interest for some LAPD users is achieving the most uniform axial profile possible. We explore this problem in the context of mirrors. The gas puff duration is known to be a large actuator for controlling density and temperature and so is explored as a way of shaping the axial profile. We predict discharges with a flat 1 kG field with the probe in the center. The discharge voltage was set at 110 (a reasonable, middle value) with the run set flag=\texttt{DR2} and top gas puffing=off. The inferred effect of gas puff duration on the axial gradient and axial gradient scale length can be seen in Fig. \ref{fig:axial-grad_gas-puff.pdf}. Care was taken to handle the aleatoric (independent) uncertainty separately from the axially-dependent epistemic uncertainty. As seen in the figure, the mean axial gradient decreases when the gas puff duration is shortened. The gradient scale length also increases, so the mean gradient is not decreasing simply because the bulk $I_\text{sat}$ is decreasing. This result suggests that the gas puff duration may be a useful actuator to consider when planning future experiments. 

These applicability of these results are somewhat muted because the gas puff duration was not chosen randomly in the training discharges. 
Given this lack of data diversity, the accuracy and applicability of this study must be interpreted cautiously. When a model is trained on \emph{all} data available (using the cross-validated test set MSE as a guide for error), which includes the 20 ms gas puff case, the predicted gradient scale length decreases uniformly across the duration scan by 1 meter. The fact that the trend remains intact when an additional, randomized intermediate gas puff case is added gives confidence in the predictions of the model despite the lack of data diversity.

\section{Optimizing profiles}
\label{sec:optimization}

One particular issue seen in LAPD plasmas is sharp axial density and temperature gradients. Some experiments require flat gradients, such as Alfv\'en wave propagation studies. We explore optimizing the axial $I_\text{sat}$ variation as an approximation to this problem. 
In addition, in this case the optimization problem is used as a tool to evaluate the quality of the learned model. This is a very demanding task because the trends inferred by the model along all inputs must simultaneously be accurate. Constraints on this optimization further increase the difficulty of the problem. Success in optimization provides strong evidence that the model has inferred relevant trends in predicting $I_\text{sat}$.
We quantify the uniformity of the axial profile by taking the standard deviation of $I_\text{sat}$ of 11 points along the z-axis ($x,y=0$). The required LAPD state for attaining the most uniform axial profile can be found by finding the minimum of this standard deviation with respect to the LAPD control parameters and flags:
\begin{equation}
	\text{Inputs} = \operatorname*{arg\,min}_{\text{Inputs} \neq z} \operatorname*{sd}_{z}(I_\text{sat}|_{x=0})
\end{equation}
The largest axial variation can likewise be found by finding the maximum. The model inputs used for this optimization can be found in Table \ref{tab:axial_optimization_inputs}. 

\begin{table}
	\small
	\centering
	\caption{Machine inputs and actuators for model inference}
	\label{tab:axial_optimization_inputs}
	\begin{tabular}{l l l l}
		Input or actuator & Range & Step & Count \\
		\hline
		Source field & 500 G to 2000 G & 250 G & 7 \\
		Mirror field & 250 G to 1500 G & 250 G & 6 \\
		Midplane field & 250 G to 1500 G & 250 G & 6 \\
		Gas puff voltage & 70 V to 90 V & 5 V & 5 \\
		Discharge voltage & 70 V to 150 V & 10 V & 9 \\
		Gas puff duration & 5 ms to 38 ms & 8.25 ms & 5\\
		Probe x positions & -50 cm to 50 cm & 2 cm & 51 \\
		Probe y positions & 0 cm & -- & -- \\
		Probe z positions & 640 cm to 1140 cm & 50 cm & 11 \\
		Probe angle & 0 rad & -- & --\\
		Run set flag & off and on & 1 & 2 \\
		Top gas puff flag & off and on & 1 & 2\\
		
	\end{tabular}
\end{table}

For this optimization we use an ensemble of five $\beta$-NLL-loss models with weight decay $\lambda=0$. The $\lambda=0$ model is used because it appears to give the most useful uncertainty estimate (seen in Fig. \ref{fig:extrapolation-profile-var_two}). The optimal machine actuator states are found by feeding a grid of inputs into the neural network. This variance estimate is not well-calibrated: the error of the predictions on the test set falls far outside the predicted uncertainty. However, this uncertainty can be used in a relative way: when the model is predicting far outside its training range, the predicted variance is much larger. The ranges of inputs into this model are seen in Table. \ref{tab:axial_optimization_inputs}. These inputs yield 127,234,800 different machine states (times five models) which takes 151 seconds to process on an RTX 3090 ($\approx 4.2$ million forward passes per second) when implemented in a naive way. The number of forward passes can be reduced by a factor of 51 if the x value is set to 0 cm. Note that gradient-based methods can be used for search because the network is differentiable everywhere but this network and parameter space is sufficiently small that a comprehensive search is computationally tractable.

Like any optimization method, the results may be pathologically optimal. In this scenario, the unconstrained minimal axial variation is found when the $I_\text{sat}$ is only around 1 mA/mm$^2$, which is quite small and corresponds to $1\text{-}2\times 10^{12}$ cm$^{-3}$ depending on Te. The inputs corresponding to this optimum are given in the second column of Table \ref{tab:axial_optimization_results}. This density range is below what is required or useful for many studies in the LAPD. 

\begin{figure*}
	\centering
	\includegraphics[width=360pt]{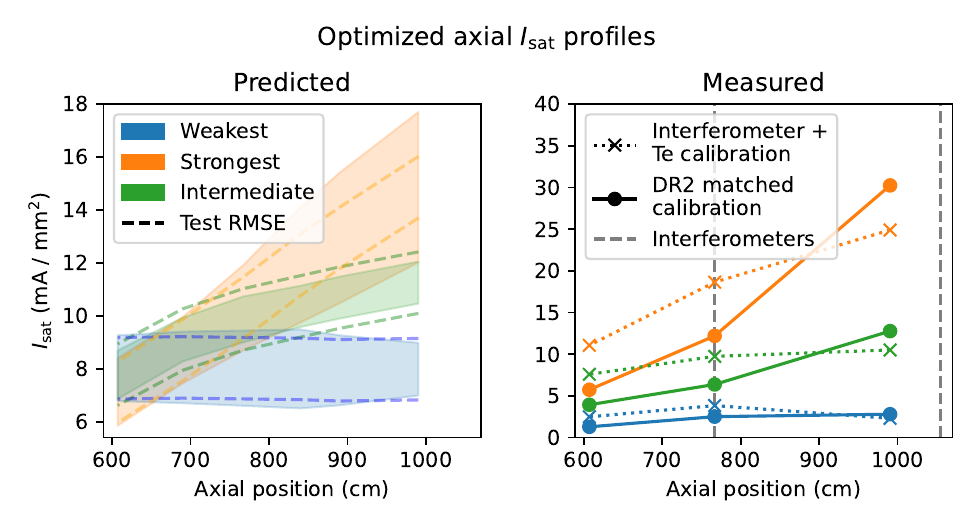}
	\caption[size=12]{\label{fig:axial-var_prediction-vs-measurement}Axial profiles, predicted and measured, for the optimized weakest (blue), intermediate (green), and strongest (orange) cases. a. The shaded region covers the mean prediction $\pm$ one standard deviation, and the dashed lines are $\pm$ the median cross-validation RMSE values. b. The measured $I_\text{sat}$ values are calibrated to \texttt{DR2} run 10 (solid lines), or using triple probe Te measurements on the probe and linearly extrapolating the interferometer measurements (dotted lines). The absolute values disagree between the predicted and measured values, but axial trends are consistent with the optimization.}
\end{figure*}

Since many physics studies require higher densities, we constrain the mean axial $I_\text{sat}$ value to be greater than 7.5 mA/mm$^{2}$ (roughly $0.5\text{-}2 \times 10^{13} \text{ cm}^{-3}$). The ``run set flag'' is set to ``on'' for cases to be validated (bolded in Table \ref{tab:axial_optimization_results}) because we wish to keep the turbopumps on to represent typical LAPD operating conditions. In addition the ``top gas puff flag'' was set to `off' to minimize the complexity of operating the fueling system on followup dataruns and experiments. Turning the top gas puff valve on is predicted to decrease the average $I_\text{sat}$ by $-2$ mA/mm$^2$ for strongly varying profiles, but otherwise the shapes are very similar.
The inputs corresponding to the maximum and minimum axial variation under these constraints can be seen in columns 3 and 4 of Table \ref{tab:axial_optimization_results}. For contrast we also consider what machine settings would lead to the greatest axial variation. The results of both of these optimizations can be seen in Fig. \ref{fig:axial-var_prediction-vs-measurement}. The optimizations yield profiles that have the largest $I_\text{sat}$ values closest to the cathode, which is expected.

The prediction for an intermediate axial variation case is also seen in Fig. \ref{fig:axial-var_prediction-vs-measurement}. The intermediate case was chosen as somewhere around half way between the strongest and weakest case with a round index number (15000, in this case). The parameters for intermediate case are also enumerated in Table \ref{tab:axial_optimization_results}. 

\begin{table*}
	\small
	\centering
	\caption{Machine inputs and actuators for optimized axial profiles}
	\label{tab:axial_optimization_results}
	\begin{tabular}{p{1.8 in} | p{0.75 in} p{0.75 in} p{0.75 in} p{0.8 in}}
		Input or actuator & Weakest & \textbf{Weakest} & \textbf{Strongest} & Intermediate \\
		$I_\text{sat}$ constraint (mA/mm$^2$) & $I_\text{sat} = $ any & $I_\text{sat}>7.5$ & $I_\text{sat}>7.5$ & $I_\text{sat}>7.5$\\
		\hline
		Source field      & 750 G   & 1000 G   & 500 G 	& 2000 G \\
		Mirror field      & 1000 G  & 750 G    & 500 G 	& 1250 G \\
		Midplane field    & 250 G   & 250 G    & 1500 G & 750 G \\
		Gas puff voltage  & 70 V    & 75 V     & 90 V 	& 90 V \\
		Discharge voltage & 130 V   & 150 V    & 150 V 	& 120 V \\
		Gas puff duration & 5 ms    & 5 ms     & 38 ms 	& 38 ms \\
		Run set flag      & on      & on       & on 	& on \\
		Top gas puff flag & on      & off      & off 	& off \\
	\end{tabular}
\end{table*}

The predicted configurations with the run set flag on and top gas puff flag off (bolded in Table \ref{tab:axial_optimization_results} were then applied on the LAPD. The data collected, compared with the predictions can be seen in Fig. \ref{fig:axial-var_prediction-vs-measurement}.

For the optimized axial profiles, the absolute value of the $I_\text{sat}$ predictions compared to measurement do not agree. All of the predicted profiles have overlapping predictions (within the predicted error) at the region furthest from the cathode, but the measured values do not show that behavior. Although the mean $I_\text{sat}$ value was constrained to be greater than 7.5 mA/mm$^2$, the measured mean was 2 mA/mm$^2$ for the weakest case.

The important result is that the optimized LAPD settings, when implemented on the machine, do yield profiles with strong, intermediate, and weak axial variation. Although the minimum-$I_\text{sat}$ constraint was violated for the case of weakest axial variation case, this optimization would nonetheless be very useful for creating axial profiles of the desired shape. 

There are three contributing factors to the mismatch of the ML-predicted values and the real measured values. First, the condition of the machine, such as the cathode emissivity or temperature or the downstream neutral pressure, are unquantified and cannot be compensated for in data preprocessing or in the model itself. Second, the calibration of the Langmuir probes could differ substantially between runs. The probes in the training data run sets (\texttt{DR1} and \texttt{DR2}) were well-calibrated to each other within the run set, but were not absolutely calibrated. The probes used for verifying the optimization were not calibrated. A rough calibration was performed by linearly extrapolating interferometer measurements and using triple probes (dotted lines on the right panel in Fig. \ref{fig:axial-var_prediction-vs-measurement}). A configuration identical to \texttt{DR2} run 10 was also measured to simultaneously correct cathode condition and probe calibration (solid lines on the right panel in Fig. \ref{fig:axial-var_prediction-vs-measurement}). Langmuir probe calibration is discussed further in Appendix \ref{app:calib}. Third, the original dataset may not have sufficient diversity to make accurate predictions on such a constrained optimization problem.

If this optimization were performed using the dataset instead of the model, the constrained search would encompass just 10084 shots out of the 131550 shots total in the training dataset, or around 7.7\%. Including the on-axis constraint reduces the number of shots down to 303 (270 in the training set), or 0.23\% of all shots in the dataset. We conclude that this optimization of an arbitrary objective function, as done here, would be intractable using traditional, non-machine learning techniques because orders of magnitude more dataruns would need to be collected. 

Optimization requires correctly learning the trends of all inputs and how they interact. In addition, as seen from the shot statistics, the model was trained on very few shots in the constrained input and output space. These two factors -- the need for the model to learn all trends and the constrained search space -- combine to make an incredibly difficult task that functions as a benchmark for the model. These factors considered, it is not surprising that the model incorrectly predicts the absolute value. The uncertainty predicted by the model, though not well-calibrated, was nonetheless very large compared to the median test set RMSE. The model did predict the trends correctly, however; the optimized, measured profiles were strong, intermediate, or weak.

We did not check to see if the predicted optima were actually true optima: an approximation of the local derivative using a finite-difference technique would require much more run time on the LAPD than was available. 

\section{Discussion}
\label{sec:discussion}

\subsection{Key achievements}

To the authors' knowledge, this work is one of the first instances machine learning has been used to infer specific trends and optimize profiles in magnetized plasmas. Three examples of trend inference were shown in this paper: influence of magnetic geometry on plasma width, changes in the axial and radial profiles with changing discharge voltage, and the relationship of gas puff duration with axial gradient scale length. In addition, the axial profile was optimized by minimizing (or maximizing) the axial standard deviation. There is no other way of simultaneously uncovering many trends or finding optima without using an ML model trained on a diverse dataset. Traditionally, such studies would require extensive scans over grids to map the parameter space, but here it was accomplished with a relatively small amount of data.

The trends inferred in this work, such as changing discharge voltages, gas puff durations, or mirror fields, would traditionally require a grid scan (varying one parameter at a time) in LAPD settings space. Here instead we are able to extract any trend covered by the training set with only a minimal amount of machine configurations sampled. Both data collection runs lacked absolute $I_\text{sat}$ calibration and had potential differences in cathode condition. Despite these issues the model learned trends that were exploited via optimization. 

In addition, this work demonstrates uncertainty quantification broken down into epistemic and aleatoric components by using ensembles and a negative-log likelihood loss. This uncertainty estimate is useful in gauging relative certainty between different predictions of the model which increases confidence in the predictions of the model. In general, the total uncertainty predicted by the model increases when predictions are made outside the bounds of the training data. 

Fundamentally, this model can predict $I_\text{sat}$ with uncertainty at any point in space covered by the training data. No other model currently exists that can perform this prediction. Traditionally, this capability would be possible only with a detailed theoretical study. 

\subsection{Current limitations}

This study would be dramatically improved by collecting more, diverse data. Only 44 of the 67 dataruns in this dataset had randomly sampled LAPD machine settings which is very small compared to the over 60,000 possible combinations. In addition, there are many other settings or parameters that were not changed in this study, such as gas puff timings, gas puff valve asymmetries, wall/limiter biasing, cathode heater settings, operation of the north cathode, and so on. The bounds of the inputs were also conservative; all settings in this study could be pushed higher or lower with a small amount of risk to LAPD operations. In addition, the placement of the probes could be further varied and placed outside the mirror cell, which would provide a more complete picture of LAPD plasmas, particularly axial effects.

Probe calibrations differed between the two training run sets (\texttt{DR1} and \texttt{DR2}) -- and a flag was introduced to distinguish between them -- but despite this deficiency combining the two run sets was shown to be advantageous for model performance. The condition of the cathode (e.g., electron emissivity and uniformity) also has a large impact on the measured $I_\text{sat}$. The improved model performance with the flag suggests that inconsistencies between dataruns could be compensated for using latent variables if a generative modeling approach is to be taken. At the very least, this model provides a way to benchmark these differences in machine state.

Concerning the model, hyperparameter tuning could be performed. In this study a few extra percent in MSE performance is not meaningful considering the limited dataset. Instead, we focused on the trends and insights that can be extracted from this model rather than simple predictive accuracy. There may also be regimes in hyperparameter space where the uncertainty is better calibrated (perhaps using early stopping). Uncertainty estimation is important, even if the absolute uncertainty is not well-calibrated because it can provide a useful relative estimate as shown in this paper. 

Trends such as the dependence of axial gradient on the gas puff duration (fig. \ref{fig:axial-grad_gas-puff.pdf}) or the effect discharge voltage on x-z profiles (fig. \ref{fig:discharge_voltage_effect}), although intuitive, remain unverified. Verification of these trends would increase confidence of model predictions when setting LAPD parameters in future experiments.

\subsection{Future directions}

The neural network architecture used here can readily scale to additional inputs and outputs; including time-series signals is the obvious next step. Integration of multiple diagnostics -- perhaps starting with individual models before combining them -- could enable inference of plasma parameters throughout the device volume. For example, combining triple probe electron temperature measurements with existing $I_\text{sat}$ data would allow density predictions anywhere in the plasma. This capability could enable in-situ diagnostic cross-calibration (e.g., the Thomson scattering density measurement) and prediction of higher-order distribution moments like particle flux. The model could be further enhanced by incorporating physics constraints such as boundary conditions (e.g., zero $I_\text{sat}$ at the machine wall) or symmetries. 

The problem presented here -- learning time-averaged $I_\text{sat}$ trends -- is fairly simple and required a relatively simple model. As demonstrated in this work, ML provides a way to explore regions of parameter space quickly and efficiently. Most physics studies on plasma devices (and fusion devices) are dedicated to a single particular problem, use grid scans, and are not useful to other experiments or campaigns. This work shows a way of using data and trends uncovered from other experimental studies. This work also demonstrates that random exploration can be a useful tool: the increased diversity of the aggregated data will generally benefit an ML model whether the experimenter discovers something new or not.

\section{Conclusion}
\label{sec:conclusion}

We demonstrate the first randomized experiments in a magnetized plasma experiment to train a neural network model. This learned model was then used to infer trends when changing field configuration, discharge voltage, or gas puff duration. This model was also used to optimize axial variation of $I_\text{sat}$ as measured by the standard deviation, which was validated in later experiments despite poor absolute error. 

We strongly advocate that all ML-based analyses in plasma and fusion research should be validated  and used to gain insight by inferring trends, as demonstrated here. This validation step is crucial for ensuring that ML models capture physically meaningful relationships and the insights provided may provide direction for future research. We hope this is the first step towards automating plasma science.

\section{Acknowledgements}

The authors would like to thank Prof. Walter Gekelman, Prof. Christoph Niemann, Dr. Shreekrishna Tripathi, Dr. Steve Vincena, Dr. Pat Pribyl, Tom Look, Yhoshua Wug, Dr. Lukas Rovige, and Dr. Jia Han for insightful discussions and experimental support.

This work was performed at the Basic Plasma Science Facility, which is a DOE Office of Science, FES collaborative user facility and is funded by DOE (DE-FC02-07ER54918) and the National Science Foundation (NSF-PHY 1036140).

\appendix

\section{The open dataset and repository} 
\label{app:repo}
All the code to perform the ML portion of this study is available at \url{https://doi.org/10.5281/zenodo.15007853}. The training datasets are also available in that repository in the \texttt{datasets} directory. Additional data are available upon request. The repository also contains additional training details and the notebooks for generating the plots in this document. The code and dataset are licensed under Creative Commons Attribution Share Alike 4.0 International.

\section{Data bias \label{sec:app_bias}}

\begin{figure*}
	\centering
	\includegraphics[width=\textwidth]{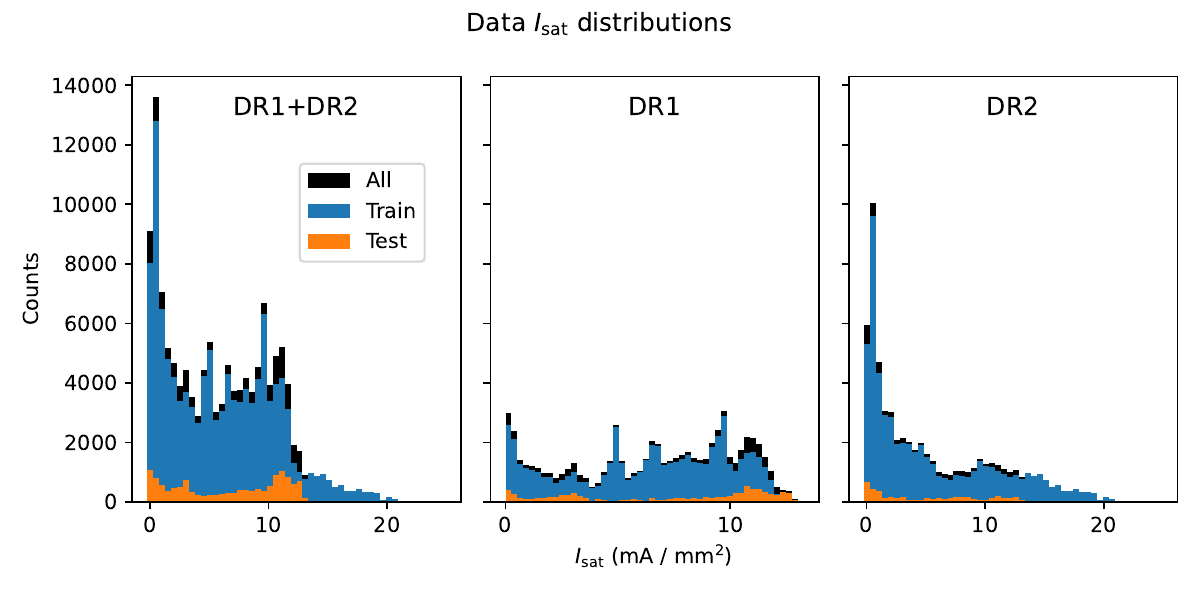}
	\caption[size=12]{\label{fig:PP1_02_isat_distribution}Distribution of $I_\text{sat}$ signals. \texttt{DR1} appears to have a more uniform distribution than \texttt{DR2} does. Combining the two datasets results in many $I_\text{sat}$ examples near 0 mA/mm$^2$ and a sharp decrease in number of examples above 10 mA/mm$^2$. From these histograms we expect or model to be biased towards fitting lower $I_\text{sat}$ values better, and to perform badly in cases with very high $I_\text{sat}$ values.}
\end{figure*}

Despite the best efforts to randomize the machine configuration, imbalance in the dataset will be present because of the relatively small amount of samples for the given actuator space. The distribution of $I_\text{sat}$ signals can be seen in Fig. \ref{fig:PP1_02_isat_distribution}. The $I_\text{sat}$ distribution is clearly different for \texttt{DR1} and \texttt{DR2}, with \texttt{DR1} having a much flatter distribution. These distributions imply that if the model is constrained to sample from \texttt{DR2} via the run set flag, then the model is expected to predict a lower $I_\text{sat}$ value in general. When predicting from the model in general, performance will likely be worse for $I_\text{sat}$ values $\gtrsim 11$ mA/mm$^2$. 

\begin{table*}
\small
	\centering
	\caption{Data breakdown by class and dataset (percent)}
	\label{tab:data_frac}
	\begin{tabular}{lrrr|lrrr|lrrr}
		\multicolumn{4}{c|}{B source (G)} & \multicolumn{4}{c|}{B mirror (G)} & \multicolumn{4}{c}{B midplane (G)}\\
		\hline \hline
		& Train & Test & All && Train & Test & All && Train & Test & All \\
		500 & 4.77 & 0 & 4.29 & 250 & 4.30 & 8.41 & 4.72 & 250 & 8.25 & 21.01 & 9.55 \\
		750 & 3.34 & 12.61 & 4.29 & 500 & 30.49 & 8.41 & 28.23 & 500 & 43.80 & 8.41 & 40.19 \\
		1000 & 43.13 & 78.99 & 46.78 & 750 & 6.68 & 16.81 & 7.72 & 750 & 6.62 & 52.19 & 11.27 \\
		1250 & 12.59 & 0 & 11.30 & 1000 & 28.85 & 57.97 & 31.82 & 1000 & 26.36 & 5.78 & 24.26 \\
		1500 & 19.23 & 0 & 17.27 & 1250 & 3.34 & 4.20 & 3.43 & 1250 & 9.24 & 0 & 8.30 \\
		1750 & 1.91 & 0 & 1.71 & 1500 & 26.34 & 4.20 & 24.08 & 1500 & 5.73 & 12.61 & 6.43 \\
		2000 & 15.03 & 8.41 & 14.35 & & & & & & & & \\
		\\
		\multicolumn{4}{c|}{Gas puff voltage (V)} & \multicolumn{4}{c|}{Discharge voltage (V)} & \multicolumn{4}{c}{Axial probe position (cm)} \\
		\hline \hline
		70 & 12.11 & 16.81 & 12.59  & 70 & 12.22 & 8.41 & 11.83    & 639 & 12.48 & 8.41 & 12.06 \\
		75 & 6.68 & 0 & 6.00     & 80 & 5.25 & 0 & 4.72      & 828 & 17.07 & 36.28 & 19.03 \\
		80 & 11.46 & 8.41 & 11.15   & 90 & 2.86 & 8.41 & 3.43      & 859 & 12.48 & 8.41 & 12.06  \\
		82 & 41.49 & 57.97 & 43.17  & 100 & 3.34 & 8.41 & 3.86     & 895 & 33.01 & 30.10 & 32.71 \\
		85 & 14.13 & 0 & 12.69   & 110 & 8.77 & 0 & 7.87     & 1017 & 12.48 & 8.41 & 12.06 \\
		90 & 14.13 & 16.81 & 14.40  & 112 & 20.62 & 0 & 18.52   & 1145 & 12.48 & 8.41 & 12.06 \\
                      & & & & 120 & 3.82 & 8.41 & 4.29     &                       & & & \\
                      & & & & 130 & 0.95 & 0 & 0.86     &                       & & & \\
                      & & & & 140 & 2.86 & 8.41 & 3.43     &                       & & & \\
                      & & & & 150 & 39.30 & 57.97 & 41.20  &                       & & & \\
		\\
		\multicolumn{4}{c|}{Gas puff duration (ms)} & \multicolumn{4}{c}{Vertical probe position (cm)}\\
		\cline{0-7} \cline{0-7}
		$38$ & 94.27 & 91.59 & 94.00 & $\approx 0$ & 36.26 & 46.08 & 37.26 & \\
		$<38$ & 5.73 & 8.41 & 6.00    & $\neq 0$ & 63.74 & 53.92 & 62.74    & \\
		\multicolumn{12}{l}{}
	\end{tabular}
\end{table*}

The distribution of the selected machine settings for all the dataruns is enumerated in Table \ref{tab:data_frac}. Despite the randomization of the settings of 44 dataruns, the distribution is often uneven. This unevenness is exacerbated in the test set because that is a selection of 6 out of 67 dataruns. The remaining 23 non-random dataruns also contribute to the imbalance. For example, a source field of 1 kG and discharge voltage of 112 show up disproportionately in the dataset because data were collected at those settings while other equipment was being adjusted or calibrated.

\section{Data and training pipeline validation}

\begin{figure}
	\includegraphics[width=\columnwidth]{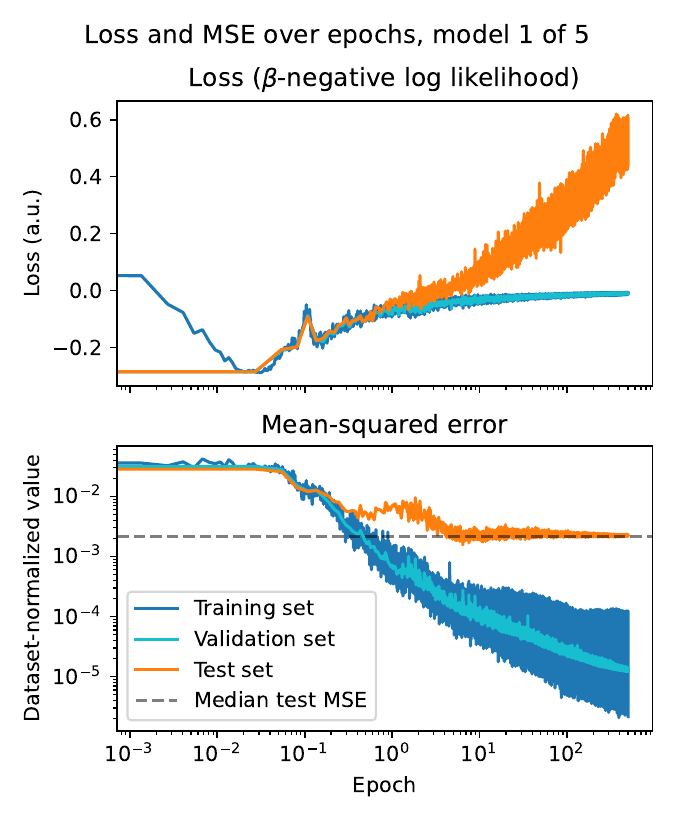}
	\caption[size=12]{\label{fig:beta-NLL_loss-mse}The loss and MSE for the training, validation, and test sets over the entire training duration of 500 epochs. The inclusion of the $\beta$ term in the loss function -- interpreted as a per-example learning rate -- makes the loss function no longer interpretable in simple terms. The mean-squared error benefits from longer training for all sets.}
\end{figure}

Multiple models were trained with varying depths and widths to verify that training loss decreases with increased model capacity. Doubling the layer width from 512 to 1024 moderately decreases the training loss; doubling the depth of the network from 4 to 8 layers has a larger impact. Increasing the width further to 2048 and depth to 12 layers has a dramatic impact on training loss, so this model and dataset are behaving nominally.
The loss function for one of the NNs in the ensemble is seen in Fig. \ref{fig:beta-NLL_loss-mse}. The MSE decreases monotonically for the training and validation set, but does not for the test set. The loss function can no longer be interpreted as a log-likelihood because of the effective per-example learning rate set by the $\beta$ term in the loss (eq. \ref{eq:loss_beta-NLL}). Note that early stopping (at around 8 epochs) would improve the test set loss, but the MSE would still be several factors higher than after 500 epochs. Early stopping was not explored in this study.

\section{Effect of $I_\text{sat}$ calibration \label{app:calib}}

\begin{figure}
	\includegraphics[width=\columnwidth]{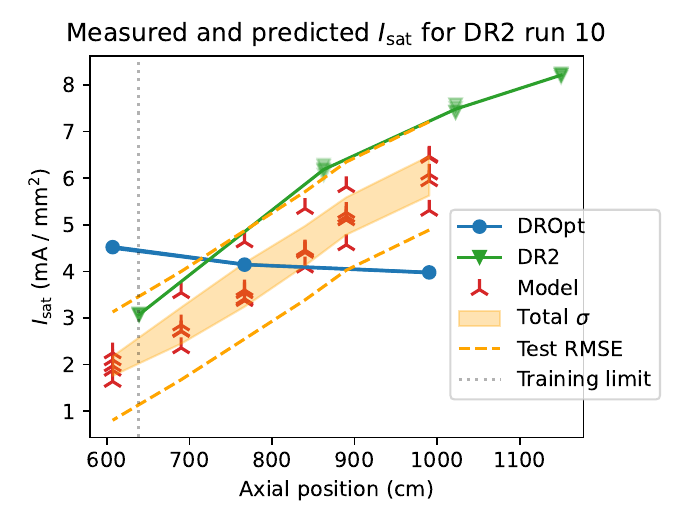}
	\caption[size=12]{\label{fig:DR2-10_LHS-30_valdiation}Comparison of original \texttt{DR2} profiles with the profiles from the optimization dataset (\texttt{DROpt}) for the same machine configuration. The $I_\text{sat}$ values in the \texttt{DROpt} dataset are not calibrated in this plot, indicating significant variation in probe calibration in this \texttt{DROpt} dataset.}
\end{figure}

The Langmuir probes did not seem to be behaving correctly when the optimization validation data were taken. The probes showed an \emph{increasing} $I_\text{sat}$ profile when moving further from the cathode in the lowest gas puff condition, which is in direct disagreement with previous measurements and intuition. An example of this discrepancy can be seen in Fig. \ref{fig:DR2-10_LHS-30_valdiation}, where a run from the original testing set (specifically \texttt{DR2} run 10) is duplicated. The probes for the validation run can be either corrected for by assuming the 5 ms gas puff run has a flat axial profile, or normalizing the probes to the \texttt{DR2} run 10 axial profile. Calibrating the probes using the \texttt{DR2} run 10 reference was the best way to go because it corrects for both probe discrepancies as well as changes in the condition (or emissivity) of the main cathode. 

\bibliography{bibliography}

%aipnum4-2.bst 2019-01-14 (MD) hand-edited version of apsrev4-1.bst
\begin{thebibliography}{30}%
\makeatletter
\providecommand \@ifxundefined [1]{%
 \@ifx{#1\undefined}
}%
\providecommand \@ifnum [1]{%
 \ifnum #1\expandafter \@firstoftwo
 \else \expandafter \@secondoftwo
 \fi
}%
\providecommand \@ifx [1]{%
 \ifx #1\expandafter \@firstoftwo
 \else \expandafter \@secondoftwo
 \fi
}%
\providecommand \natexlab [1]{#1}%
\providecommand \enquote  [1]{``#1''}%
\providecommand \bibnamefont  [1]{#1}%
\providecommand \bibfnamefont [1]{#1}%
\providecommand \citenamefont [1]{#1}%
\providecommand \href@noop [0]{\@secondoftwo}%
\providecommand \href [0]{\begingroup \@sanitize@url \@href}%
\providecommand \@href[1]{\@@startlink{#1}\@@href}%
\providecommand \@@href[1]{\endgroup#1\@@endlink}%
\providecommand \@sanitize@url [0]{\catcode `\\12\catcode `\$12\catcode `\&12\catcode `\#12\catcode `\^12\catcode `\_12\catcode `\%12\relax}%
\providecommand \@@startlink[1]{}%
\providecommand \@@endlink[0]{}%
\providecommand \url  [0]{\begingroup\@sanitize@url \@url }%
\providecommand \@url [1]{\endgroup\@href {#1}{\urlprefix }}%
\providecommand \urlprefix  [0]{URL }%
\providecommand \Eprint [0]{\href }%
\providecommand \doibase [0]{https://doi.org/}%
\providecommand \selectlanguage [0]{\@gobble}%
\providecommand \bibinfo  [0]{\@secondoftwo}%
\providecommand \bibfield  [0]{\@secondoftwo}%
\providecommand \translation [1]{[#1]}%
\providecommand \BibitemOpen [0]{}%
\providecommand \bibitemStop [0]{}%
\providecommand \bibitemNoStop [0]{.\EOS\space}%
\providecommand \EOS [0]{\spacefactor3000\relax}%
\providecommand \BibitemShut  [1]{\csname bibitem#1\endcsname}%
\let\auto@bib@innerbib\@empty
%</preamble>
\bibitem [{\citenamefont {Abbate}, \citenamefont {Conlin},\ and\ \citenamefont {Kolemen}(2021)}]{abbate_data-driven_2021}%
  \BibitemOpen
  \bibfield  {author} {\bibinfo {author} {\bibfnamefont {J.}~\bibnamefont {Abbate}}, \bibinfo {author} {\bibfnamefont {R.}~\bibnamefont {Conlin}},\ and\ \bibinfo {author} {\bibfnamefont {E.}~\bibnamefont {Kolemen}},\ }\href {https://doi.org/10.1088/1741-4326/abe08d} {\bibfield  {journal} {\bibinfo  {journal} {Nuclear Fusion}\ }\textbf {\bibinfo {volume} {61}},\ \bibinfo {pages} {046027} (\bibinfo {year} {2021})}\BibitemShut {NoStop}%
\bibitem [{\citenamefont {Jalalvand}\ \emph {et~al.}(2022)\citenamefont {Jalalvand}, \citenamefont {Abbate}, \citenamefont {Conlin}, \citenamefont {Verdoolaege},\ and\ \citenamefont {Kolemen}}]{jalalvand_real-time_2022}%
  \BibitemOpen
  \bibfield  {author} {\bibinfo {author} {\bibfnamefont {A.}~\bibnamefont {Jalalvand}}, \bibinfo {author} {\bibfnamefont {J.}~\bibnamefont {Abbate}}, \bibinfo {author} {\bibfnamefont {R.}~\bibnamefont {Conlin}}, \bibinfo {author} {\bibfnamefont {G.}~\bibnamefont {Verdoolaege}},\ and\ \bibinfo {author} {\bibfnamefont {E.}~\bibnamefont {Kolemen}},\ }\href {https://doi.org/10.1109/TNNLS.2021.3085504} {\bibfield  {journal} {\bibinfo  {journal} {IEEE Transactions on Neural Networks and Learning Systems}\ }\textbf {\bibinfo {volume} {33}},\ \bibinfo {pages} {2630} (\bibinfo {year} {2022})}\BibitemShut {NoStop}%
\bibitem [{\citenamefont {Char}\ \emph {et~al.}(2024)\citenamefont {Char}, \citenamefont {Chung}, \citenamefont {Abbate}, \citenamefont {Kolemen},\ and\ \citenamefont {Schneider}}]{char_full_2024}%
  \BibitemOpen
  \bibfield  {author} {\bibinfo {author} {\bibfnamefont {I.}~\bibnamefont {Char}}, \bibinfo {author} {\bibfnamefont {Y.}~\bibnamefont {Chung}}, \bibinfo {author} {\bibfnamefont {J.}~\bibnamefont {Abbate}}, \bibinfo {author} {\bibfnamefont {E.}~\bibnamefont {Kolemen}},\ and\ \bibinfo {author} {\bibfnamefont {J.}~\bibnamefont {Schneider}},\ }\href {http://arxiv.org/abs/2404.12416} {\enquote {\bibinfo {title} {Full {Shot} {Predictions} for the {DIII}-{D} {Tokamak} via {Deep} {Recurrent} {Networks}},}\ } (\bibinfo {year} {2024}),\ \bibinfo {note} {arXiv:2404.12416 [physics]}\BibitemShut {NoStop}%
\bibitem [{\citenamefont {Wan}\ \emph {et~al.}(2022)\citenamefont {Wan}, \citenamefont {Yu}, \citenamefont {Pau}, \citenamefont {Liu},\ and\ \citenamefont {Li}}]{wan_east_2022}%
  \BibitemOpen
  \bibfield  {author} {\bibinfo {author} {\bibfnamefont {C.}~\bibnamefont {Wan}}, \bibinfo {author} {\bibfnamefont {Z.}~\bibnamefont {Yu}}, \bibinfo {author} {\bibfnamefont {A.}~\bibnamefont {Pau}}, \bibinfo {author} {\bibfnamefont {X.}~\bibnamefont {Liu}},\ and\ \bibinfo {author} {\bibfnamefont {J.}~\bibnamefont {Li}},\ }\href {https://doi.org/10.1088/1741-4326/ac9c1a} {\bibfield  {journal} {\bibinfo  {journal} {Nuclear Fusion}\ }\textbf {\bibinfo {volume} {62}},\ \bibinfo {pages} {126060} (\bibinfo {year} {2022})}\BibitemShut {NoStop}%
\bibitem [{\citenamefont {Seo}\ \emph {et~al.}(2021)\citenamefont {Seo}, \citenamefont {Na}, \citenamefont {Kim}, \citenamefont {Lee}, \citenamefont {Park}, \citenamefont {Park},\ and\ \citenamefont {Lee}}]{seo_feedforward_2021}%
  \BibitemOpen
  \bibfield  {author} {\bibinfo {author} {\bibfnamefont {J.}~\bibnamefont {Seo}}, \bibinfo {author} {\bibfnamefont {Y.-S.}\ \bibnamefont {Na}}, \bibinfo {author} {\bibfnamefont {B.}~\bibnamefont {Kim}}, \bibinfo {author} {\bibfnamefont {C.}~\bibnamefont {Lee}}, \bibinfo {author} {\bibfnamefont {M.}~\bibnamefont {Park}}, \bibinfo {author} {\bibfnamefont {S.}~\bibnamefont {Park}},\ and\ \bibinfo {author} {\bibfnamefont {Y.}~\bibnamefont {Lee}},\ }\href {https://doi.org/doi:10.1088/1741-4326/ac121b} {\bibfield  {journal} {\bibinfo  {journal} {Nuclear Fusion}\ }\textbf {\bibinfo {volume} {61}},\ \bibinfo {pages} {106010} (\bibinfo {year} {2021})}\BibitemShut {NoStop}%
\bibitem [{\citenamefont {Seo}\ \emph {et~al.}(2022)\citenamefont {Seo}, \citenamefont {Na}, \citenamefont {Kim}, \citenamefont {Lee}, \citenamefont {Park}, \citenamefont {Park},\ and\ \citenamefont {Lee}}]{seo_development_2022}%
  \BibitemOpen
  \bibfield  {author} {\bibinfo {author} {\bibfnamefont {J.}~\bibnamefont {Seo}}, \bibinfo {author} {\bibfnamefont {Y.-S.}\ \bibnamefont {Na}}, \bibinfo {author} {\bibfnamefont {B.}~\bibnamefont {Kim}}, \bibinfo {author} {\bibfnamefont {C.}~\bibnamefont {Lee}}, \bibinfo {author} {\bibfnamefont {M.}~\bibnamefont {Park}}, \bibinfo {author} {\bibfnamefont {S.}~\bibnamefont {Park}},\ and\ \bibinfo {author} {\bibfnamefont {Y.}~\bibnamefont {Lee}},\ }\href {https://doi.org/10.1088/1741-4326/ac79be} {\bibfield  {journal} {\bibinfo  {journal} {Nuclear Fusion}\ }\textbf {\bibinfo {volume} {62}},\ \bibinfo {pages} {086049} (\bibinfo {year} {2022})}\BibitemShut {NoStop}%
\bibitem [{\citenamefont {Fu}\ \emph {et~al.}(2020)\citenamefont {Fu}, \citenamefont {Eldon}, \citenamefont {Erickson}, \citenamefont {Kleijwegt}, \citenamefont {Lupin-Jimenez}, \citenamefont {Boyer}, \citenamefont {Eidietis}, \citenamefont {Barbour}, \citenamefont {Izacard},\ and\ \citenamefont {Kolemen}}]{fu_machine_2020}%
  \BibitemOpen
  \bibfield  {author} {\bibinfo {author} {\bibfnamefont {Y.}~\bibnamefont {Fu}}, \bibinfo {author} {\bibfnamefont {D.}~\bibnamefont {Eldon}}, \bibinfo {author} {\bibfnamefont {K.}~\bibnamefont {Erickson}}, \bibinfo {author} {\bibfnamefont {K.}~\bibnamefont {Kleijwegt}}, \bibinfo {author} {\bibfnamefont {L.}~\bibnamefont {Lupin-Jimenez}}, \bibinfo {author} {\bibfnamefont {M.~D.}\ \bibnamefont {Boyer}}, \bibinfo {author} {\bibfnamefont {N.}~\bibnamefont {Eidietis}}, \bibinfo {author} {\bibfnamefont {N.}~\bibnamefont {Barbour}}, \bibinfo {author} {\bibfnamefont {O.}~\bibnamefont {Izacard}},\ and\ \bibinfo {author} {\bibfnamefont {E.}~\bibnamefont {Kolemen}},\ }\href {https://doi.org/10.1063/1.5125581} {\bibfield  {journal} {\bibinfo  {journal} {Physics of Plasmas}\ }\textbf {\bibinfo {volume} {27}},\ \bibinfo {pages} {022501} (\bibinfo {year} {2020})}\BibitemShut {NoStop}%
\bibitem [{\citenamefont {Dong}\ \emph {et~al.}(2021)\citenamefont {Dong}, \citenamefont {Li}, \citenamefont {Ding}, \citenamefont {Zhang}, \citenamefont {Wang}, \citenamefont {Li}, \citenamefont {Yan}, \citenamefont {Shen}, \citenamefont {He}, \citenamefont {Ren},\ and\ \citenamefont {Xia}}]{dong_machine_2021}%
  \BibitemOpen
  \bibfield  {author} {\bibinfo {author} {\bibfnamefont {J.}~\bibnamefont {Dong}}, \bibinfo {author} {\bibfnamefont {J.}~\bibnamefont {Li}}, \bibinfo {author} {\bibfnamefont {Y.}~\bibnamefont {Ding}}, \bibinfo {author} {\bibfnamefont {X.}~\bibnamefont {Zhang}}, \bibinfo {author} {\bibfnamefont {N.}~\bibnamefont {Wang}}, \bibinfo {author} {\bibfnamefont {D.}~\bibnamefont {Li}}, \bibinfo {author} {\bibfnamefont {W.}~\bibnamefont {Yan}}, \bibinfo {author} {\bibfnamefont {C.}~\bibnamefont {Shen}}, \bibinfo {author} {\bibfnamefont {Y.}~\bibnamefont {He}}, \bibinfo {author} {\bibfnamefont {X.}~\bibnamefont {Ren}},\ and\ \bibinfo {author} {\bibfnamefont {D.}~\bibnamefont {Xia}},\ }\href {https://doi.org/10.1088/2058-6272/ac0685} {\bibfield  {journal} {\bibinfo  {journal} {Plasma Science and Technology}\ }\textbf {\bibinfo {volume} {23}},\ \bibinfo {pages} {085101} (\bibinfo {year} {2021})}\BibitemShut {NoStop}%
\bibitem [{\citenamefont {Kates-Harbeck}, \citenamefont {Svyatkovskiy},\ and\ \citenamefont {Tang}(2019)}]{kates-harbeck_predicting_2019}%
  \BibitemOpen
  \bibfield  {author} {\bibinfo {author} {\bibfnamefont {J.}~\bibnamefont {Kates-Harbeck}}, \bibinfo {author} {\bibfnamefont {A.}~\bibnamefont {Svyatkovskiy}},\ and\ \bibinfo {author} {\bibfnamefont {W.}~\bibnamefont {Tang}},\ }\href {https://doi.org/10.1038/s41586-019-1116-4} {\bibfield  {journal} {\bibinfo  {journal} {Nature}\ }\textbf {\bibinfo {volume} {568}},\ \bibinfo {pages} {526} (\bibinfo {year} {2019})}\BibitemShut {NoStop}%
\bibitem [{\citenamefont {Rea}\ \emph {et~al.}(2019)\citenamefont {Rea}, \citenamefont {Montes}, \citenamefont {Erickson}, \citenamefont {Granetz},\ and\ \citenamefont {Tinguely}}]{rea_real-time_2019}%
  \BibitemOpen
  \bibfield  {author} {\bibinfo {author} {\bibfnamefont {C.}~\bibnamefont {Rea}}, \bibinfo {author} {\bibfnamefont {K.}~\bibnamefont {Montes}}, \bibinfo {author} {\bibfnamefont {K.}~\bibnamefont {Erickson}}, \bibinfo {author} {\bibfnamefont {R.}~\bibnamefont {Granetz}},\ and\ \bibinfo {author} {\bibfnamefont {R.}~\bibnamefont {Tinguely}},\ }\href {https://doi.org/10.1088/1741-4326/ab28bf} {\bibfield  {journal} {\bibinfo  {journal} {Nuclear Fusion}\ }\textbf {\bibinfo {volume} {59}},\ \bibinfo {pages} {096016} (\bibinfo {year} {2019})}\BibitemShut {NoStop}%
\bibitem [{\citenamefont {Vega}\ \emph {et~al.}(2022)\citenamefont {Vega}, \citenamefont {Murari}, \citenamefont {Dormido-Canto}, \citenamefont {Rattá}, \citenamefont {Gelfusa},\ and\ \citenamefont {{JET Contributors}}}]{vega_disruption_2022}%
  \BibitemOpen
  \bibfield  {author} {\bibinfo {author} {\bibfnamefont {J.}~\bibnamefont {Vega}}, \bibinfo {author} {\bibfnamefont {A.}~\bibnamefont {Murari}}, \bibinfo {author} {\bibfnamefont {S.}~\bibnamefont {Dormido-Canto}}, \bibinfo {author} {\bibfnamefont {G.~A.}\ \bibnamefont {Rattá}}, \bibinfo {author} {\bibfnamefont {M.}~\bibnamefont {Gelfusa}},\ and\ \bibinfo {author} {\bibnamefont {{JET Contributors}}},\ }\href {https://doi.org/10.1038/s41567-022-01602-2} {\bibfield  {journal} {\bibinfo  {journal} {Nature Physics}\ }\textbf {\bibinfo {volume} {18}},\ \bibinfo {pages} {741} (\bibinfo {year} {2022})}\BibitemShut {NoStop}%
\bibitem [{\citenamefont {Seo}\ \emph {et~al.}(2024)\citenamefont {Seo}, \citenamefont {Kim}, \citenamefont {Jalalvand}, \citenamefont {Conlin}, \citenamefont {Rothstein}, \citenamefont {Abbate}, \citenamefont {Erickson}, \citenamefont {Wai}, \citenamefont {Shousha},\ and\ \citenamefont {Kolemen}}]{seo_avoiding_2024}%
  \BibitemOpen
  \bibfield  {author} {\bibinfo {author} {\bibfnamefont {J.}~\bibnamefont {Seo}}, \bibinfo {author} {\bibfnamefont {S.}~\bibnamefont {Kim}}, \bibinfo {author} {\bibfnamefont {A.}~\bibnamefont {Jalalvand}}, \bibinfo {author} {\bibfnamefont {R.}~\bibnamefont {Conlin}}, \bibinfo {author} {\bibfnamefont {A.}~\bibnamefont {Rothstein}}, \bibinfo {author} {\bibfnamefont {J.}~\bibnamefont {Abbate}}, \bibinfo {author} {\bibfnamefont {K.}~\bibnamefont {Erickson}}, \bibinfo {author} {\bibfnamefont {J.}~\bibnamefont {Wai}}, \bibinfo {author} {\bibfnamefont {R.}~\bibnamefont {Shousha}},\ and\ \bibinfo {author} {\bibfnamefont {E.}~\bibnamefont {Kolemen}},\ }\href {https://doi.org/10.1038/s41586-024-07024-9} {\bibfield  {journal} {\bibinfo  {journal} {Nature}\ }\textbf {\bibinfo {volume} {626}},\ \bibinfo {pages} {746} (\bibinfo {year} {2024})}\BibitemShut {NoStop}%
\bibitem [{\citenamefont {Murari}\ \emph {et~al.}(2020)\citenamefont {Murari}, \citenamefont {Peluso}, \citenamefont {Lungaroni}, \citenamefont {Rossi}, \citenamefont {Gelfusa},\ and\ \citenamefont {{JET Contributors}}}]{murari_investigating_2020}%
  \BibitemOpen
  \bibfield  {author} {\bibinfo {author} {\bibfnamefont {A.}~\bibnamefont {Murari}}, \bibinfo {author} {\bibfnamefont {E.}~\bibnamefont {Peluso}}, \bibinfo {author} {\bibfnamefont {M.}~\bibnamefont {Lungaroni}}, \bibinfo {author} {\bibfnamefont {R.}~\bibnamefont {Rossi}}, \bibinfo {author} {\bibfnamefont {M.}~\bibnamefont {Gelfusa}},\ and\ \bibinfo {author} {\bibnamefont {{JET Contributors}}},\ }\href {https://doi.org/10.3390/app10196683} {\bibfield  {journal} {\bibinfo  {journal} {Applied Sciences}\ }\textbf {\bibinfo {volume} {10}},\ \bibinfo {pages} {6683} (\bibinfo {year} {2020})}\BibitemShut {NoStop}%
\bibitem [{\citenamefont {Maris}\ \emph {et~al.}(2024)\citenamefont {Maris}, \citenamefont {Rea}, \citenamefont {Pau}, \citenamefont {Hu}, \citenamefont {Xiao}, \citenamefont {Granetz}, \citenamefont {Marmar}, \citenamefont {team}, \citenamefont {team}, \citenamefont {team}, \citenamefont {team}, \citenamefont {team},\ and\ \citenamefont {team}}]{maris_correlation_2024}%
  \BibitemOpen
  \bibfield  {author} {\bibinfo {author} {\bibfnamefont {A.}~\bibnamefont {Maris}}, \bibinfo {author} {\bibfnamefont {C.}~\bibnamefont {Rea}}, \bibinfo {author} {\bibfnamefont {A.}~\bibnamefont {Pau}}, \bibinfo {author} {\bibfnamefont {W.}~\bibnamefont {Hu}}, \bibinfo {author} {\bibfnamefont {B.}~\bibnamefont {Xiao}}, \bibinfo {author} {\bibfnamefont {R.}~\bibnamefont {Granetz}}, \bibinfo {author} {\bibfnamefont {E.}~\bibnamefont {Marmar}}, \bibinfo {author} {\bibfnamefont {t.~E. T.~E.}\ \bibnamefont {team}}, \bibinfo {author} {\bibfnamefont {t.~A. C.-M.}\ \bibnamefont {team}}, \bibinfo {author} {\bibfnamefont {t.~A.~U.}\ \bibnamefont {team}}, \bibinfo {author} {\bibfnamefont {t.~D.-D.}\ \bibnamefont {team}}, \bibinfo {author} {\bibfnamefont {t.~E.}\ \bibnamefont {team}},\ and\ \bibinfo {author} {\bibfnamefont {t.~T.}\ \bibnamefont {team}},\ }\href {http://arxiv.org/abs/2406.18442} {\enquote {\bibinfo {title} {Correlation of the {L}-mode density limit with edge collisionality},}\ } (\bibinfo {year} {2024}),\ \bibinfo {note} {arXiv:2406.18442 [physics]}\BibitemShut {NoStop}%
\bibitem [{\citenamefont {Pavone}\ \emph {et~al.}(2023)\citenamefont {Pavone}, \citenamefont {Merlo}, \citenamefont {Kwak},\ and\ \citenamefont {Svensson}}]{pavone_machine_2023}%
  \BibitemOpen
  \bibfield  {author} {\bibinfo {author} {\bibfnamefont {A.}~\bibnamefont {Pavone}}, \bibinfo {author} {\bibfnamefont {A.}~\bibnamefont {Merlo}}, \bibinfo {author} {\bibfnamefont {S.}~\bibnamefont {Kwak}},\ and\ \bibinfo {author} {\bibfnamefont {J.}~\bibnamefont {Svensson}},\ }\href {https://doi.org/10.1088/1361-6587/acc60f} {\bibfield  {journal} {\bibinfo  {journal} {Plasma Physics and Controlled Fusion}\ }\textbf {\bibinfo {volume} {65}},\ \bibinfo {pages} {053001} (\bibinfo {year} {2023})}\BibitemShut {NoStop}%
\bibitem [{\citenamefont {Döpp}\ \emph {et~al.}(2023)\citenamefont {Döpp}, \citenamefont {Eberle}, \citenamefont {Howard}, \citenamefont {Irshad}, \citenamefont {Lin},\ and\ \citenamefont {Streeter}}]{dopp_data-driven_2023}%
  \BibitemOpen
  \bibfield  {author} {\bibinfo {author} {\bibfnamefont {A.}~\bibnamefont {Döpp}}, \bibinfo {author} {\bibfnamefont {C.}~\bibnamefont {Eberle}}, \bibinfo {author} {\bibfnamefont {S.}~\bibnamefont {Howard}}, \bibinfo {author} {\bibfnamefont {F.}~\bibnamefont {Irshad}}, \bibinfo {author} {\bibfnamefont {J.}~\bibnamefont {Lin}},\ and\ \bibinfo {author} {\bibfnamefont {M.}~\bibnamefont {Streeter}},\ }\href {https://doi.org/10.1017/hpl.2023.47} {\bibfield  {journal} {\bibinfo  {journal} {High Power Laser Science and Engineering}\ }\textbf {\bibinfo {volume} {11}},\ \bibinfo {pages} {e55} (\bibinfo {year} {2023})}\BibitemShut {NoStop}%
\bibitem [{\citenamefont {Anirudh}\ \emph {et~al.}(2023)\citenamefont {Anirudh}, \citenamefont {Archibald}, \citenamefont {Asif}, \citenamefont {Becker}, \citenamefont {Benkadda}, \citenamefont {Bremer}, \citenamefont {Budé}, \citenamefont {Chang}, \citenamefont {Chen}, \citenamefont {Churchill}, \citenamefont {Citrin}, \citenamefont {Gaffney}, \citenamefont {Gainaru}, \citenamefont {Gekelman}, \citenamefont {Gibbs}, \citenamefont {Hamaguchi}, \citenamefont {Hill}, \citenamefont {Humbird}, \citenamefont {Jalas}, \citenamefont {Kawaguchi}, \citenamefont {Kim}, \citenamefont {Kirchen}, \citenamefont {Klasky}, \citenamefont {Kline}, \citenamefont {Krushelnick}, \citenamefont {Kustowski}, \citenamefont {Lapenta}, \citenamefont {Li}, \citenamefont {Ma}, \citenamefont {Mason}, \citenamefont {Mesbah}, \citenamefont {Michoski}, \citenamefont {Munson}, \citenamefont {Murakami}, \citenamefont {Najm}, \citenamefont {Olofsson}, \citenamefont {Park}, \citenamefont {Peterson}, \citenamefont {Probst}, \citenamefont {Pugmire}, \citenamefont {Sammuli}, \citenamefont {Sawlani}, \citenamefont {Scheinker}, \citenamefont {Schissel}, \citenamefont {Shalloo}, \citenamefont {Shinagawa}, \citenamefont {Seong}, \citenamefont {Spears}, \citenamefont {Tennyson}, \citenamefont {Thiagarajan}, \citenamefont {Ticoş}, \citenamefont {Trieschmann}, \citenamefont {Dijk}, \citenamefont {Essen}, \citenamefont {Ventzek}, \citenamefont {Wang}, \citenamefont {Wang}, \citenamefont {Wang}, \citenamefont {Wende}, \citenamefont {Xu}, \citenamefont {Yamada}, \citenamefont {Yokoyama},\ and\ \citenamefont {Zhang}}]{anirudh_2022_2023}%
  \BibitemOpen
  \bibfield  {author} {\bibinfo {author} {\bibfnamefont {R.}~\bibnamefont {Anirudh}}, \bibinfo {author} {\bibfnamefont {R.}~\bibnamefont {Archibald}}, \bibinfo {author} {\bibfnamefont {M.~S.}\ \bibnamefont {Asif}}, \bibinfo {author} {\bibfnamefont {M.~M.}\ \bibnamefont {Becker}}, \bibinfo {author} {\bibfnamefont {S.}~\bibnamefont {Benkadda}}, \bibinfo {author} {\bibfnamefont {P.-T.}\ \bibnamefont {Bremer}}, \bibinfo {author} {\bibfnamefont {R.~H.~S.}\ \bibnamefont {Budé}}, \bibinfo {author} {\bibfnamefont {C.~S.}\ \bibnamefont {Chang}}, \bibinfo {author} {\bibfnamefont {L.}~\bibnamefont {Chen}}, \bibinfo {author} {\bibfnamefont {R.~M.}\ \bibnamefont {Churchill}}, \bibinfo {author} {\bibfnamefont {J.}~\bibnamefont {Citrin}}, \bibinfo {author} {\bibfnamefont {J.~A.}\ \bibnamefont {Gaffney}}, \bibinfo {author} {\bibfnamefont {A.}~\bibnamefont {Gainaru}}, \bibinfo {author} {\bibfnamefont {W.}~\bibnamefont {Gekelman}}, \bibinfo {author} {\bibfnamefont {T.}~\bibnamefont {Gibbs}}, \bibinfo {author} {\bibfnamefont {S.}~\bibnamefont {Hamaguchi}}, \bibinfo {author} {\bibfnamefont {C.}~\bibnamefont {Hill}}, \bibinfo {author} {\bibfnamefont {K.}~\bibnamefont {Humbird}}, \bibinfo {author} {\bibfnamefont {S.}~\bibnamefont {Jalas}}, \bibinfo {author} {\bibfnamefont {S.}~\bibnamefont {Kawaguchi}}, \bibinfo {author} {\bibfnamefont {G.-H.}\ \bibnamefont {Kim}}, \bibinfo {author} {\bibfnamefont {M.}~\bibnamefont {Kirchen}}, \bibinfo {author} {\bibfnamefont {S.}~\bibnamefont {Klasky}}, \bibinfo {author} {\bibfnamefont {J.~L.}\ \bibnamefont {Kline}}, \bibinfo {author} {\bibfnamefont {K.}~\bibnamefont {Krushelnick}}, \bibinfo {author} {\bibfnamefont {B.}~\bibnamefont {Kustowski}}, \bibinfo {author} {\bibfnamefont {G.}~\bibnamefont {Lapenta}}, \bibinfo {author} {\bibfnamefont {W.}~\bibnamefont {Li}}, \bibinfo {author} {\bibfnamefont {T.}~\bibnamefont {Ma}}, \bibinfo {author} {\bibfnamefont {N.~J.}\ \bibnamefont {Mason}}, \bibinfo {author} {\bibfnamefont {A.}~\bibnamefont {Mesbah}}, \bibinfo {author} {\bibfnamefont {C.}~\bibnamefont
  {Michoski}}, \bibinfo {author} {\bibfnamefont {T.}~\bibnamefont {Munson}}, \bibinfo {author} {\bibfnamefont {I.}~\bibnamefont {Murakami}}, \bibinfo {author} {\bibfnamefont {H.~N.}\ \bibnamefont {Najm}}, \bibinfo {author} {\bibfnamefont {K.~E.~J.}\ \bibnamefont {Olofsson}}, \bibinfo {author} {\bibfnamefont {S.}~\bibnamefont {Park}}, \bibinfo {author} {\bibfnamefont {J.~L.}\ \bibnamefont {Peterson}}, \bibinfo {author} {\bibfnamefont {M.}~\bibnamefont {Probst}}, \bibinfo {author} {\bibfnamefont {D.}~\bibnamefont {Pugmire}}, \bibinfo {author} {\bibfnamefont {B.}~\bibnamefont {Sammuli}}, \bibinfo {author} {\bibfnamefont {K.}~\bibnamefont {Sawlani}}, \bibinfo {author} {\bibfnamefont {A.}~\bibnamefont {Scheinker}}, \bibinfo {author} {\bibfnamefont {D.~P.}\ \bibnamefont {Schissel}}, \bibinfo {author} {\bibfnamefont {R.~J.}\ \bibnamefont {Shalloo}}, \bibinfo {author} {\bibfnamefont {J.}~\bibnamefont {Shinagawa}}, \bibinfo {author} {\bibfnamefont {J.}~\bibnamefont {Seong}}, \bibinfo {author} {\bibfnamefont {B.~K.}\ \bibnamefont {Spears}}, \bibinfo {author} {\bibfnamefont {J.}~\bibnamefont {Tennyson}}, \bibinfo {author} {\bibfnamefont {J.}~\bibnamefont {Thiagarajan}}, \bibinfo {author} {\bibfnamefont {C.~M.}\ \bibnamefont {Ticoş}}, \bibinfo {author} {\bibfnamefont {J.}~\bibnamefont {Trieschmann}}, \bibinfo {author} {\bibfnamefont {J.~V.}\ \bibnamefont {Dijk}}, \bibinfo {author} {\bibfnamefont {B.~V.}\ \bibnamefont {Essen}}, \bibinfo {author} {\bibfnamefont {P.}~\bibnamefont {Ventzek}}, \bibinfo {author} {\bibfnamefont {H.}~\bibnamefont {Wang}}, \bibinfo {author} {\bibfnamefont {J.~T.~L.}\ \bibnamefont {Wang}}, \bibinfo {author} {\bibfnamefont {Z.}~\bibnamefont {Wang}}, \bibinfo {author} {\bibfnamefont {K.}~\bibnamefont {Wende}}, \bibinfo {author} {\bibfnamefont {X.}~\bibnamefont {Xu}}, \bibinfo {author} {\bibfnamefont {H.}~\bibnamefont {Yamada}}, \bibinfo {author} {\bibfnamefont {T.}~\bibnamefont {Yokoyama}},\ and\ \bibinfo {author} {\bibfnamefont {X.}~\bibnamefont {Zhang}},\ }\href
  {https://doi.org/10.1109/TPS.2023.3268170} {\bibfield  {journal} {\bibinfo  {journal} {IEEE Transactions on Plasma Science}\ }\textbf {\bibinfo {volume} {51}},\ \bibinfo {pages} {1750} (\bibinfo {year} {2023})}\BibitemShut {NoStop}%
\bibitem [{\citenamefont {Daly}\ \emph {et~al.}(2023)\citenamefont {Daly}, \citenamefont {Fieldsend}, \citenamefont {Hassall},\ and\ \citenamefont {Tabor}}]{daly_data-driven_2023}%
  \BibitemOpen
  \bibfield  {author} {\bibinfo {author} {\bibfnamefont {G.~A.}\ \bibnamefont {Daly}}, \bibinfo {author} {\bibfnamefont {J.~E.}\ \bibnamefont {Fieldsend}}, \bibinfo {author} {\bibfnamefont {G.}~\bibnamefont {Hassall}},\ and\ \bibinfo {author} {\bibfnamefont {G.~R.}\ \bibnamefont {Tabor}},\ }\href {https://doi.org/10.1088/2632-2153/aced7f} {\bibfield  {journal} {\bibinfo  {journal} {Machine Learning: Science and Technology}\ }\textbf {\bibinfo {volume} {4}},\ \bibinfo {pages} {035035} (\bibinfo {year} {2023})}\BibitemShut {NoStop}%
\bibitem [{\citenamefont {Travis}(2025)}]{phil_travis_2025_15007853}%
  \BibitemOpen
  \bibfield  {author} {\bibinfo {author} {\bibfnamefont {P.}~\bibnamefont {Travis}},\ }\href {https://doi.org/10.5281/zenodo.15007853} {\enquote {\bibinfo {title} {physicistphil/lapd-isat-predict: 2025-3-11},}\ } (\bibinfo {year} {2025}),\ \bibinfo {note} {doi:10.5281/zenodo.15007853}\BibitemShut {NoStop}%
\bibitem [{\citenamefont {Roach}\ \emph {et~al.}(2008)\citenamefont {Roach}, \citenamefont {Walters}, \citenamefont {Budny}, \citenamefont {Imbeaux}, \citenamefont {Fredian}, \citenamefont {Greenwald}, \citenamefont {Stillerman}, \citenamefont {Alexander}, \citenamefont {Carlsson}, \citenamefont {Cary}, \citenamefont {Ryter}, \citenamefont {Stober}, \citenamefont {Gohil}, \citenamefont {Greenfield}, \citenamefont {Murakami}, \citenamefont {Bracco}, \citenamefont {Esposito}, \citenamefont {Romanelli}, \citenamefont {Parail}, \citenamefont {Stubberfield}, \citenamefont {Voitsekhovitch}, \citenamefont {Brickley}, \citenamefont {Field}, \citenamefont {Sakamoto}, \citenamefont {Fujita}, \citenamefont {Fukuda}, \citenamefont {Hayashi}, \citenamefont {Hogeweij}, \citenamefont {Chudnovskiy}, \citenamefont {Kinerva}, \citenamefont {Kessel}, \citenamefont {Aniel}, \citenamefont {Hoang}, \citenamefont {Ongena}, \citenamefont {Doyle}, \citenamefont {Houlberg}, \citenamefont {Polevoi}, \citenamefont {{ITPA Confinement Database and Modelling Topical Group}},\ and\ \citenamefont {{ITPA Transport Physics Topical Group}}}]{roach_2008_2008}%
  \BibitemOpen
  \bibfield  {author} {\bibinfo {author} {\bibfnamefont {C.}~\bibnamefont {Roach}}, \bibinfo {author} {\bibfnamefont {M.}~\bibnamefont {Walters}}, \bibinfo {author} {\bibfnamefont {R.}~\bibnamefont {Budny}}, \bibinfo {author} {\bibfnamefont {F.}~\bibnamefont {Imbeaux}}, \bibinfo {author} {\bibfnamefont {T.}~\bibnamefont {Fredian}}, \bibinfo {author} {\bibfnamefont {M.}~\bibnamefont {Greenwald}}, \bibinfo {author} {\bibfnamefont {J.}~\bibnamefont {Stillerman}}, \bibinfo {author} {\bibfnamefont {D.}~\bibnamefont {Alexander}}, \bibinfo {author} {\bibfnamefont {J.}~\bibnamefont {Carlsson}}, \bibinfo {author} {\bibfnamefont {J.}~\bibnamefont {Cary}}, \bibinfo {author} {\bibfnamefont {F.}~\bibnamefont {Ryter}}, \bibinfo {author} {\bibfnamefont {J.}~\bibnamefont {Stober}}, \bibinfo {author} {\bibfnamefont {P.}~\bibnamefont {Gohil}}, \bibinfo {author} {\bibfnamefont {C.}~\bibnamefont {Greenfield}}, \bibinfo {author} {\bibfnamefont {M.}~\bibnamefont {Murakami}}, \bibinfo {author} {\bibfnamefont {G.}~\bibnamefont {Bracco}}, \bibinfo {author} {\bibfnamefont {B.}~\bibnamefont {Esposito}}, \bibinfo {author} {\bibfnamefont {M.}~\bibnamefont {Romanelli}}, \bibinfo {author} {\bibfnamefont {V.}~\bibnamefont {Parail}}, \bibinfo {author} {\bibfnamefont {P.}~\bibnamefont {Stubberfield}}, \bibinfo {author} {\bibfnamefont {I.}~\bibnamefont {Voitsekhovitch}}, \bibinfo {author} {\bibfnamefont {C.}~\bibnamefont {Brickley}}, \bibinfo {author} {\bibfnamefont {A.}~\bibnamefont {Field}}, \bibinfo {author} {\bibfnamefont {Y.}~\bibnamefont {Sakamoto}}, \bibinfo {author} {\bibfnamefont {T.}~\bibnamefont {Fujita}}, \bibinfo {author} {\bibfnamefont {T.}~\bibnamefont {Fukuda}}, \bibinfo {author} {\bibfnamefont {N.}~\bibnamefont {Hayashi}}, \bibinfo {author} {\bibfnamefont {G.}~\bibnamefont {Hogeweij}}, \bibinfo {author} {\bibfnamefont {A.}~\bibnamefont {Chudnovskiy}}, \bibinfo {author} {\bibfnamefont {N.}~\bibnamefont {Kinerva}}, \bibinfo {author} {\bibfnamefont {C.}~\bibnamefont {Kessel}}, \bibinfo {author} {\bibfnamefont {T.}~\bibnamefont
  {Aniel}}, \bibinfo {author} {\bibfnamefont {G.}~\bibnamefont {Hoang}}, \bibinfo {author} {\bibfnamefont {J.}~\bibnamefont {Ongena}}, \bibinfo {author} {\bibfnamefont {E.}~\bibnamefont {Doyle}}, \bibinfo {author} {\bibfnamefont {W.}~\bibnamefont {Houlberg}}, \bibinfo {author} {\bibfnamefont {A.}~\bibnamefont {Polevoi}}, \bibinfo {author} {\bibnamefont {{ITPA Confinement Database and Modelling Topical Group}}},\ and\ \bibinfo {author} {\bibnamefont {{ITPA Transport Physics Topical Group}}},\ }\href {https://doi.org/10.1088/0029-5515/48/12/125001} {\bibfield  {journal} {\bibinfo  {journal} {Nuclear Fusion}\ }\textbf {\bibinfo {volume} {48}},\ \bibinfo {pages} {125001} (\bibinfo {year} {2008})}\BibitemShut {NoStop}%
\bibitem [{\citenamefont {Jackson}\ \emph {et~al.}(2024)\citenamefont {Jackson}, \citenamefont {Khan}, \citenamefont {Cummings}, \citenamefont {Hodson}, \citenamefont {De~Witt}, \citenamefont {Pamela}, \citenamefont {Akers},\ and\ \citenamefont {Thiyagalingam}}]{jackson_fair-mast_2024}%
  \BibitemOpen
  \bibfield  {author} {\bibinfo {author} {\bibfnamefont {S.}~\bibnamefont {Jackson}}, \bibinfo {author} {\bibfnamefont {S.}~\bibnamefont {Khan}}, \bibinfo {author} {\bibfnamefont {N.}~\bibnamefont {Cummings}}, \bibinfo {author} {\bibfnamefont {J.}~\bibnamefont {Hodson}}, \bibinfo {author} {\bibfnamefont {S.}~\bibnamefont {De~Witt}}, \bibinfo {author} {\bibfnamefont {S.}~\bibnamefont {Pamela}}, \bibinfo {author} {\bibfnamefont {R.}~\bibnamefont {Akers}},\ and\ \bibinfo {author} {\bibfnamefont {J.}~\bibnamefont {Thiyagalingam}},\ }\href {https://doi.org/10.1016/j.softx.2024.101869} {\bibfield  {journal} {\bibinfo  {journal} {SoftwareX}\ }\textbf {\bibinfo {volume} {27}},\ \bibinfo {pages} {101869} (\bibinfo {year} {2024})}\BibitemShut {NoStop}%
\bibitem [{lhd()}]{lhd_data}%
  \BibitemOpen
  \href@noop {} {\enquote {\bibinfo {title} {Lhd experiment data repository},}\ }\bibinfo {note} {Doi:10.57451/lhd.analyzed-data}\BibitemShut {NoStop}%
\bibitem [{\citenamefont {Gekelman}\ \emph {et~al.}(2016)\citenamefont {Gekelman}, \citenamefont {Pribyl}, \citenamefont {Lucky}, \citenamefont {Drandell}, \citenamefont {Leneman}, \citenamefont {Maggs}, \citenamefont {Vincena}, \citenamefont {Van~Compernolle}, \citenamefont {Tripathi}, \citenamefont {Morales}, \citenamefont {Carter}, \citenamefont {Wang},\ and\ \citenamefont {DeHaas}}]{gekelman_upgraded_2016}%
  \BibitemOpen
  \bibfield  {author} {\bibinfo {author} {\bibfnamefont {W.}~\bibnamefont {Gekelman}}, \bibinfo {author} {\bibfnamefont {P.}~\bibnamefont {Pribyl}}, \bibinfo {author} {\bibfnamefont {Z.}~\bibnamefont {Lucky}}, \bibinfo {author} {\bibfnamefont {M.}~\bibnamefont {Drandell}}, \bibinfo {author} {\bibfnamefont {D.}~\bibnamefont {Leneman}}, \bibinfo {author} {\bibfnamefont {J.}~\bibnamefont {Maggs}}, \bibinfo {author} {\bibfnamefont {S.}~\bibnamefont {Vincena}}, \bibinfo {author} {\bibfnamefont {B.}~\bibnamefont {Van~Compernolle}}, \bibinfo {author} {\bibfnamefont {S.~K.~P.}\ \bibnamefont {Tripathi}}, \bibinfo {author} {\bibfnamefont {G.}~\bibnamefont {Morales}}, \bibinfo {author} {\bibfnamefont {T.~A.}\ \bibnamefont {Carter}}, \bibinfo {author} {\bibfnamefont {Y.}~\bibnamefont {Wang}},\ and\ \bibinfo {author} {\bibfnamefont {T.}~\bibnamefont {DeHaas}},\ }\href {https://doi.org/10.1063/1.4941079} {\bibfield  {journal} {\bibinfo  {journal} {Review of Scientific Instruments}\ }\textbf {\bibinfo {volume} {87}},\ \bibinfo {pages} {025105} (\bibinfo {year} {2016})}\BibitemShut {NoStop}%
\bibitem [{\citenamefont {Qian}\ \emph {et~al.}(2023)\citenamefont {Qian}, \citenamefont {Gekelman}, \citenamefont {Pribyl}, \citenamefont {Sketchley}, \citenamefont {Tripathi}, \citenamefont {Lucky}, \citenamefont {Drandell}, \citenamefont {Vincena}, \citenamefont {Look}, \citenamefont {Travis}, \citenamefont {Carter}, \citenamefont {Wan}, \citenamefont {Cattelan}, \citenamefont {Sabiston}, \citenamefont {Ottaviano},\ and\ \citenamefont {Wirz}}]{qian_design_2023}%
  \BibitemOpen
  \bibfield  {author} {\bibinfo {author} {\bibfnamefont {Y.}~\bibnamefont {Qian}}, \bibinfo {author} {\bibfnamefont {W.}~\bibnamefont {Gekelman}}, \bibinfo {author} {\bibfnamefont {P.}~\bibnamefont {Pribyl}}, \bibinfo {author} {\bibfnamefont {T.}~\bibnamefont {Sketchley}}, \bibinfo {author} {\bibfnamefont {S.}~\bibnamefont {Tripathi}}, \bibinfo {author} {\bibfnamefont {Z.}~\bibnamefont {Lucky}}, \bibinfo {author} {\bibfnamefont {M.}~\bibnamefont {Drandell}}, \bibinfo {author} {\bibfnamefont {S.}~\bibnamefont {Vincena}}, \bibinfo {author} {\bibfnamefont {T.}~\bibnamefont {Look}}, \bibinfo {author} {\bibfnamefont {P.}~\bibnamefont {Travis}}, \bibinfo {author} {\bibfnamefont {T.}~\bibnamefont {Carter}}, \bibinfo {author} {\bibfnamefont {G.}~\bibnamefont {Wan}}, \bibinfo {author} {\bibfnamefont {M.}~\bibnamefont {Cattelan}}, \bibinfo {author} {\bibfnamefont {G.}~\bibnamefont {Sabiston}}, \bibinfo {author} {\bibfnamefont {A.}~\bibnamefont {Ottaviano}},\ and\ \bibinfo {author} {\bibfnamefont {R.}~\bibnamefont {Wirz}},\ }\href {https://doi.org/10.1063/5.0152216} {\bibfield  {journal} {\bibinfo  {journal} {Review of Scientific Instruments}\ }\textbf {\bibinfo {volume} {94}},\ \bibinfo {pages} {085104} (\bibinfo {year} {2023})}\BibitemShut {NoStop}%
\bibitem [{\citenamefont {Seetharaman}\ \emph {et~al.}(2020)\citenamefont {Seetharaman}, \citenamefont {Wichern}, \citenamefont {Pardo},\ and\ \citenamefont {Roux}}]{seetharaman_autoclip_2020}%
  \BibitemOpen
  \bibfield  {author} {\bibinfo {author} {\bibfnamefont {P.}~\bibnamefont {Seetharaman}}, \bibinfo {author} {\bibfnamefont {G.}~\bibnamefont {Wichern}}, \bibinfo {author} {\bibfnamefont {B.}~\bibnamefont {Pardo}},\ and\ \bibinfo {author} {\bibfnamefont {J.~L.}\ \bibnamefont {Roux}},\ }\href@noop {} {\enquote {\bibinfo {title} {{AutoClip}: {Adaptive} {Gradient} {Clipping} for {Source} {Separation} {Networks}},}\ }\bibinfo {howpublished} {http://arxiv.org/abs/2007.14469} (\bibinfo {year} {2020}),\ \bibinfo {note} {arXiv:2007.14469 [cs, eess, stat]}\BibitemShut {NoStop}%
\bibitem [{\citenamefont {Nix}\ and\ \citenamefont {Weigend}(1994)}]{nix_estimating_1994}%
  \BibitemOpen
  \bibfield  {author} {\bibinfo {author} {\bibfnamefont {D.}~\bibnamefont {Nix}}\ and\ \bibinfo {author} {\bibfnamefont {A.}~\bibnamefont {Weigend}},\ }in\ \href@noop {} {\emph {\bibinfo {booktitle} {Proceedings of 1994 {IEEE} {International} {Conference} on {Neural} {Networks} ({ICNN}'94)}}}\ (\bibinfo  {publisher} {IEEE},\ \bibinfo {address} {Orlando, FL, USA},\ \bibinfo {year} {1994})\ pp.\ \bibinfo {pages} {55--60 vol.1}\BibitemShut {NoStop}%
\bibitem [{\citenamefont {Lakshminarayanan}, \citenamefont {Pritzel},\ and\ \citenamefont {Blundell}(2017)}]{lakshminarayanan_simple_2017}%
  \BibitemOpen
  \bibfield  {author} {\bibinfo {author} {\bibfnamefont {B.}~\bibnamefont {Lakshminarayanan}}, \bibinfo {author} {\bibfnamefont {A.}~\bibnamefont {Pritzel}},\ and\ \bibinfo {author} {\bibfnamefont {C.}~\bibnamefont {Blundell}},\ }\href@noop {} {\enquote {\bibinfo {title} {Simple and {Scalable} {Predictive} {Uncertainty} {Estimation} using {Deep} {Ensembles}},}\ }\bibinfo {howpublished} {http://arxiv.org/abs/1612.01474} (\bibinfo {year} {2017}),\ \bibinfo {note} {arXiv:1612.01474 [cs, stat]}\BibitemShut {NoStop}%
\bibitem [{\citenamefont {Seitzer}\ \emph {et~al.}(2022)\citenamefont {Seitzer}, \citenamefont {Tavakoli}, \citenamefont {Antic},\ and\ \citenamefont {Martius}}]{seitzer_pitfalls_2022}%
  \BibitemOpen
  \bibfield  {author} {\bibinfo {author} {\bibfnamefont {M.}~\bibnamefont {Seitzer}}, \bibinfo {author} {\bibfnamefont {A.}~\bibnamefont {Tavakoli}}, \bibinfo {author} {\bibfnamefont {D.}~\bibnamefont {Antic}},\ and\ \bibinfo {author} {\bibfnamefont {G.}~\bibnamefont {Martius}},\ }\href@noop {} {\enquote {\bibinfo {title} {On the {Pitfalls} of {Heteroscedastic} {Uncertainty} {Estimation} with {Probabilistic} {Neural} {Networks}},}\ }\bibinfo {howpublished} {http://arxiv.org/abs/2203.09168} (\bibinfo {year} {2022}),\ \bibinfo {note} {arXiv:2203.09168 [cs, stat]}\BibitemShut {NoStop}%
\bibitem [{\citenamefont {Valdenegro-Toro}\ and\ \citenamefont {Saromo}(2022)}]{valdenegro-toro_deeper_2022}%
  \BibitemOpen
  \bibfield  {author} {\bibinfo {author} {\bibfnamefont {M.}~\bibnamefont {Valdenegro-Toro}}\ and\ \bibinfo {author} {\bibfnamefont {D.}~\bibnamefont {Saromo}},\ }\href@noop {} {\  (\bibinfo {year} {2022})},\ \bibinfo {note} {arXiv:2204.09308 [cs.LG]}\BibitemShut {NoStop}%
\bibitem [{\citenamefont {Guo}\ \emph {et~al.}(2017)\citenamefont {Guo}, \citenamefont {Pleiss}, \citenamefont {Sun},\ and\ \citenamefont {Weinberger}}]{guo_calibration_2017}%
  \BibitemOpen
  \bibfield  {author} {\bibinfo {author} {\bibfnamefont {C.}~\bibnamefont {Guo}}, \bibinfo {author} {\bibfnamefont {G.}~\bibnamefont {Pleiss}}, \bibinfo {author} {\bibfnamefont {Y.}~\bibnamefont {Sun}},\ and\ \bibinfo {author} {\bibfnamefont {K.~Q.}\ \bibnamefont {Weinberger}},\ }\href@noop {} {\enquote {\bibinfo {title} {On {Calibration} of {Modern} {Neural} {Networks}},}\ }\bibinfo {howpublished} {http://arxiv.org/abs/1706.04599} (\bibinfo {year} {2017}),\ \bibinfo {note} {arXiv:1706.04599 [cs]}\BibitemShut {NoStop}%
\end{thebibliography}%

\end{document}